\PassOptionsToPackage{unicode}{hyperref}
\PassOptionsToPackage{hyphens}{url}
\PassOptionsToPackage{dvipsnames,svgnames,x11names}{xcolor}
\documentclass[
  letterpaper,
  DIV=11,
  numbers=noendperiod]{scrartcl}
\usepackage{xcolor}
\usepackage{amsmath,amssymb}
\usepackage{iftex}
\ifPDFTeX
  \usepackage[T1]{fontenc}
  \usepackage[utf8]{inputenc}
  \usepackage{textcomp} 
\else 
  \usepackage{unicode-math} 
  \defaultfontfeatures{Scale=MatchLowercase}
  \defaultfontfeatures[\rmfamily]{Ligatures=TeX,Scale=1}
\fi
\usepackage{lmodern}
\ifPDFTeX\else
\fi
\IfFileExists{upquote.sty}{\usepackage{upquote}}{}
\IfFileExists{microtype.sty}{
  \usepackage[]{microtype}
  \UseMicrotypeSet[protrusion]{basicmath} 
}{}
\makeatletter
\@ifundefined{KOMAClassName}{
  \IfFileExists{parskip.sty}{%
    \usepackage{parskip}
  }{
    \setlength{\parindent}{0pt}
    \setlength{\parskip}{6pt plus 2pt minus 1pt}}
}{
  \KOMAoptions{parskip=half}}
\makeatother
\makeatletter
\ifx\paragraph\undefined\else
  \let\oldparagraph\paragraph
  \renewcommand{\paragraph}{
    \@ifstar
      \xxxParagraphStar
      \xxxParagraphNoStar
  }
  \newcommand{\xxxParagraphStar}[1]{\oldparagraph*{#1}\mbox{}}
  \newcommand{\xxxParagraphNoStar}[1]{\oldparagraph{#1}\mbox{}}
\fi
\ifx\subparagraph\undefined\else
  \let\oldsubparagraph\subparagraph
  \renewcommand{\subparagraph}{
    \@ifstar
      \xxxSubParagraphStar
      \xxxSubParagraphNoStar
  }
  \newcommand{\xxxSubParagraphStar}[1]{\oldsubparagraph*{#1}\mbox{}}
  \newcommand{\xxxSubParagraphNoStar}[1]{\oldsubparagraph{#1}\mbox{}}
\fi
\makeatother

\usepackage{color}
\usepackage{fancyvrb}

\DefineVerbatimEnvironment{Highlighting}{Verbatim}{commandchars=\\\{\}}
\usepackage{framed}
\definecolor{shadecolor}{RGB}{241,243,245}
\newenvironment{Shaded}{\begin{snugshade}}{\end{snugshade}}

\newcommand{\AttributeTok}[1]{\textcolor[rgb]{0.40,0.45,0.13}{#1}}

\newcommand{\CommentTok}[1]{\textcolor[rgb]{0.37,0.37,0.37}{#1}}

\newcommand{\ConstantTok}[1]{\textcolor[rgb]{0.56,0.35,0.01}{#1}}

\newcommand{\DecValTok}[1]{\textcolor[rgb]{0.68,0.00,0.00}{#1}}

\newcommand{\FunctionTok}[1]{\textcolor[rgb]{0.28,0.35,0.67}{#1}}

\newcommand{\NormalTok}[1]{\textcolor[rgb]{0.00,0.23,0.31}{#1}}

\newcommand{\OtherTok}[1]{\textcolor[rgb]{0.00,0.23,0.31}{#1}}

\newcommand{\SpecialCharTok}[1]{\textcolor[rgb]{0.37,0.37,0.37}{#1}}

\newcommand{\StringTok}[1]{\textcolor[rgb]{0.13,0.47,0.30}{#1}}

\usepackage{longtable,booktabs,array}
\usepackage{calc} 
\usepackage{etoolbox}
\makeatletter
\patchcmd\longtable{\par}{\if@noskipsec\mbox{}\fi\par}{}{}
\makeatother
\IfFileExists{footnotehyper.sty}{\usepackage{footnotehyper}}{\usepackage{footnote}}
\makesavenoteenv{longtable}
\usepackage{graphicx}
\makeatletter
\newsavebox\pandoc@box
\newcommand*\pandocbounded[1]{
  \sbox\pandoc@box{#1}%
  \Gscale@div\@tempa{\textheight}{\dimexpr\ht\pandoc@box+\dp\pandoc@box\relax}%
  \Gscale@div\@tempb{\linewidth}{\wd\pandoc@box}%
  \ifdim\@tempb\p@<\@tempa\p@\let\@tempa\@tempb\fi
  \ifdim\@tempa\p@<\p@\scalebox{\@tempa}{\usebox\pandoc@box}%
  \else\usebox{\pandoc@box}%
  \fi%
}
\def\fps@figure{htbp}
\makeatother

\NewDocumentCommand\citeproctext{}{}

\makeatletter
 \let\@cite@ofmt\@firstofone
 \def\@biblabel#1{}
 \def\@cite#1#2{{#1\if@tempswa , #2\fi}}
\makeatother
\newlength{\cslhangindent}
\newlength{\csllabelwidth}
\newenvironment{CSLReferences}[2] 
 {\begin{list}{}{%
  \setlength{\itemindent}{0pt}
  \setlength{\leftmargin}{0pt}
  \setlength{\parsep}{0pt}
  \ifodd #1
   \setlength{\leftmargin}{\cslhangindent}
   \setlength{\itemindent}{-1\cslhangindent}
  \fi
  \setlength{\itemsep}{#2\baselineskip}}}
 {\end{list}}
\usepackage{calc}

\providecommand{\tightlist}{%
  \setlength{\itemsep}{0pt}\setlength{\parskip}{0pt}}

\KOMAoption{captions}{tableheading}
\makeatletter
\@ifpackageloaded{caption}{}{\usepackage{caption}}
\AtBeginDocument{%
\ifdefined\contentsname
  \renewcommand*\contentsname{Table of contents}
\else
  \newcommand\contentsname{Table of contents}
\fi
\ifdefined\listfigurename
  \renewcommand*\listfigurename{List of Figures}
\else
  \newcommand\listfigurename{List of Figures}
\fi
\ifdefined\listtablename
  \renewcommand*\listtablename{List of Tables}
\else
  \newcommand\listtablename{List of Tables}
\fi
\ifdefined\figurename
  \renewcommand*\figurename{Figure}
\else
  \newcommand\figurename{Figure}
\fi
\ifdefined\tablename
  \renewcommand*\tablename{Table}
\else
  \newcommand\tablename{Table}
\fi
}
\@ifpackageloaded{float}{}{\usepackage{float}}
\floatstyle{ruled}
\@ifundefined{c@chapter}{\newfloat{codelisting}{h}{lop}}{\newfloat{codelisting}{h}{lop}[chapter]}
\floatname{codelisting}{Listing}

\makeatother
\makeatletter
\@ifpackageloaded{caption}{}{\usepackage{caption}}
\@ifpackageloaded{subcaption}{}{\usepackage{subcaption}}
\makeatother
\usepackage{bookmark}
\IfFileExists{xurl.sty}{\usepackage{xurl}}{} 
\hypersetup{
  pdftitle={tidychangepoint: a unified framework for analyzing changepoint detection in univariate time series},
  pdfauthor={Benjamin S. Baumer; Biviana Marcela Suárez Sierra},
  pdfkeywords={time series analysis, changepoint detection, R
packages, penalty functions, genetic algorithms},
  colorlinks=true,
  linkcolor={blue},
  filecolor={Maroon},
  citecolor={Blue},
  urlcolor={Blue},
  pdfcreator={LaTeX via pandoc}}

\title{tidychangepoint: a unified framework for analyzing changepoint
detection in univariate time series}
\author{Benjamin S. Baumer \and Biviana Marcela Suárez Sierra}
\date{2025-07-09}
\begin{document}
\maketitle
\begin{abstract}
We present {tidychangepoint}, a new {R} package for changepoint
detection analysis. Most {R} packages for segmenting univariate time
series focus on providing one or two algorithms for changepoint
detection that work with a small set of models and penalized objective
functions, and all of them return a custom, nonstandard object type.
This makes comparing results across various algorithms, models, and
penalized objective functions unnecessarily difficult. {tidychangepoint}
solves this problem by wrapping functions from a variety of existing
packages and storing the results in a common S3 class called
\texttt{tidycpt}. The package then provides functionality for easily
extracting comparable numeric or graphical information from a
\texttt{tidycpt} object, all in a {tidyverse}-compliant framework.
{tidychangepoint} is versatile: it supports both deterministic
algorithms like PELT (from {changepoint}), and also flexible,
randomized, genetic algorithms (via {GA}) that---via new functionality
built into {tidychangepoint}---can be used with any compliant
model-fitting function and any penalized objective function. By bringing
all of these disparate tools together in a cohesive fashion,
{tidychangepoint} facilitates comparative analysis of changepoint
detection algorithms and models.
\end{abstract}

\section{Introduction}\label{sec-intro}

\subsection{Motivation}\label{sec-motivation}

While global temperatures continue to rise, climate scientists continue
to face baffling skepticism from some. How do we know that the increases
we observe are not just anecdotal? How do we know that the recent
pattern of rising temperatures is not simply the latest round of normal
fluctuation?

One approach would be to measure temperature over a long period of time,
and investigate the time series for the presence of \emph{changepoints}.
Changepoints indicate breaks in a time series that divide it into
mutually exclusive \emph{regions} in which the behavior of the
underlying system is the same within each region, but varies across
regions. A changepoint might represent a moment at which the system
changed in some meaningful and measurable way.

Changepoint detection analysis is the study of such changepoints and the
algorithms and models that determine them. Unfortunately, there is no
known general solution to the Changepoint Detection Problem (CDP,
defined formally in Section~\ref{sec-math}). Furthermore---and more
germaine for us---the objects returned by changepoint detection
functions from various R packages (see Section~\ref{sec-pkgs}) are all
of different types, have different structures, contain different
information, and can be used with a variety of different functions. This
heterogeneity in user interface and package architecture is widespread
across roughly a dozen other changepoint detection packages we found
(see Section~\ref{sec-pkgs}). No standardization exists.

These incompatible software interfaces make it unnecessarily difficult
to compare the performance of changepoint detection algorithms across
different packages. One essentially has to write custom code for each
implementation, even when using the same data set. Furthermore, since
there is no pre-built interface for genetic changepoint detection
algorithms, every tweak to a genetic algorithm for changepoint detection
requires custom code.

While we do not claim to offer solutions to the CDP itself,
{tidychangepoint} solves the software heterogeneity problem by providing
a unified interface with built-in support for genetic algorithms for
changepoint detection. Comparing algorithms and/or models for
changepoint detection in {tidychangepoint} can be as easy as iterating
over a list of specifications (see Section~\ref{sec-ex-tidycpt}).

\subsection{An example illustrating the status
quo}\label{an-example-illustrating-the-status-quo}

\begin{figure}[H]

\centering{

\pandocbounded{\includegraphics[keepaspectratio]{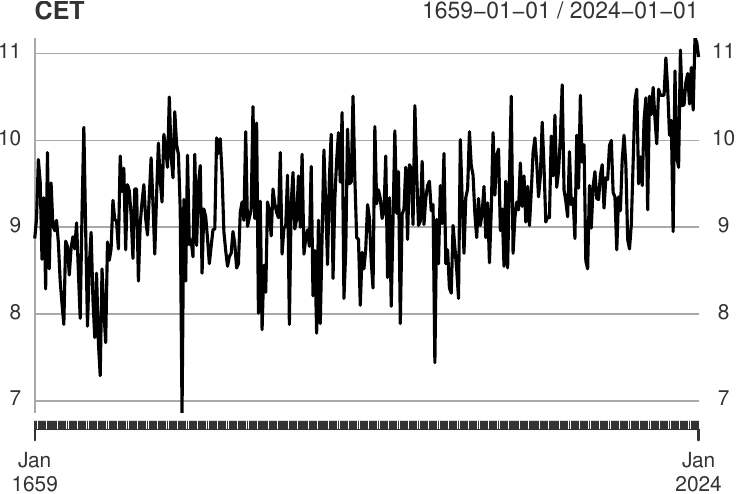}}

}

\caption{\label{fig-cet}Mean annual temperature in degrees Celsius, as
measured in Hadley Centre, England, 1659-2024. Plot created by
\texttt{plot.xts()}.}

\end{figure}

For example, the \texttt{CET} data set collected by Parker, Legg, and
Folland (1992) and repackaged in {tidychangepoint} (Baumer et al. 2025)
lists the mean annual temperature in degrees Celsius from 1659 to 2024,
as measured in Central England and shown in Figure~\ref{fig-cet}. When
did the temperature start to rise? How can we distinguish natural
variation in temperature (around a static mean) from a meaningful change
in that mean (caused by humans)?

\phantomsection\label{cell-fig-cet}
\begin{Shaded}
\begin{Highlighting}[]
\FunctionTok{library}\NormalTok{(tidychangepoint)}
\FunctionTok{plot}\NormalTok{(CET) }
\end{Highlighting}
\end{Shaded}

Suppose that you are interested in determining changepoints for the
\texttt{CET} data set, but you don't know much about the inner workings
of changepoint detection algorithms. You decide to simply fit a bunch of
off-the-shelf changepoint detection models and compare the results. Some
googling leads you to four packages: {strucchange} (Zeileis et al.
2024), {segmented} (V. M. R. Muggeo 2025), {changepoint} (Killick 2024),
and {wbs} (Baranowski and Fryzlewicz 2024).

First, you fit four models. Note that because \texttt{CET} is an
\texttt{xts} object, sometimes (but not always), you have to cast it to
a \texttt{ts} object.

\begin{Shaded}
\begin{Highlighting}[]
\NormalTok{cet\_cpt }\OtherTok{\textless{}{-}}\NormalTok{ changepoint}\SpecialCharTok{::}\FunctionTok{cpt.meanvar}\NormalTok{(}\FunctionTok{as.ts}\NormalTok{(CET))}
\NormalTok{cet\_wbs }\OtherTok{\textless{}{-}}\NormalTok{ wbs}\SpecialCharTok{::}\FunctionTok{wbs}\NormalTok{(CET)}
\NormalTok{cet\_bkp }\OtherTok{\textless{}{-}}\NormalTok{ strucchange}\SpecialCharTok{::}\FunctionTok{breakpoints}\NormalTok{(CET }\SpecialCharTok{\textasciitilde{}} \DecValTok{1}\NormalTok{)}
\NormalTok{cet\_seg }\OtherTok{\textless{}{-}}\NormalTok{ segmented}\SpecialCharTok{::}\FunctionTok{segmented}\NormalTok{(}\FunctionTok{as.ts}\NormalTok{(CET))}
\end{Highlighting}
\end{Shaded}

Next, you seek to recover the changepoints \emph{as a numeric vector}.
These are the indices of the time series (in this case, each corresponds
to a year since 1659). Note that each object (some of which are S3 and
some of which are S4) stores the changepoints in different locations,
some of which are not easy to find.

\begin{Shaded}
\begin{Highlighting}[]
\NormalTok{cet\_cpt}\SpecialCharTok{@}\NormalTok{cpts}
\end{Highlighting}
\end{Shaded}

\begin{verbatim}
[1]  55  57 267 344 347 366
\end{verbatim}

\begin{Shaded}
\begin{Highlighting}[]
\FunctionTok{sort}\NormalTok{(cet\_wbs}\SpecialCharTok{$}\NormalTok{cpt}\SpecialCharTok{$}\NormalTok{cpt.ic}\SpecialCharTok{$}\NormalTok{mbic.penalty)}
\end{Highlighting}
\end{Shaded}

\begin{verbatim}
[1]  29  43  81  82 234 330
\end{verbatim}

\begin{Shaded}
\begin{Highlighting}[]
\NormalTok{cet\_bkp}\SpecialCharTok{$}\NormalTok{breakpoints}
\end{Highlighting}
\end{Shaded}

\begin{verbatim}
[1]  55 252 312
\end{verbatim}

\begin{Shaded}
\begin{Highlighting}[]
\NormalTok{cet\_seg}\SpecialCharTok{$}\NormalTok{psi[ , }\DecValTok{2}\NormalTok{]}
\end{Highlighting}
\end{Shaded}

\begin{verbatim}
[1] 321.788
\end{verbatim}

\begin{figure}

\begin{minipage}{0.50\linewidth}

\centering{

\pandocbounded{\includegraphics[keepaspectratio]{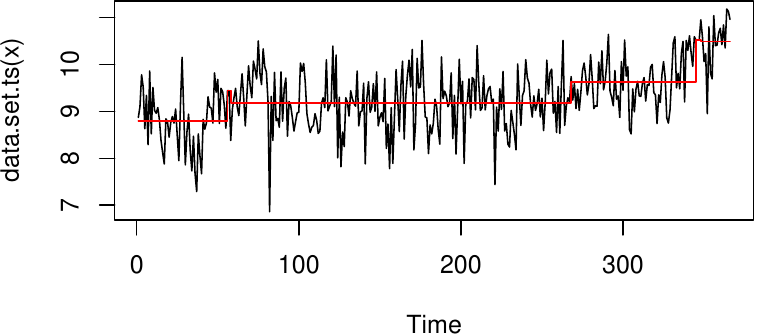}}

}

\subcaption{\label{fig-cet-old-1}PELT. Plot created by
\texttt{plot.cpt()}.}

\end{minipage}%
\begin{minipage}{0.50\linewidth}

\centering{

\pandocbounded{\includegraphics[keepaspectratio]{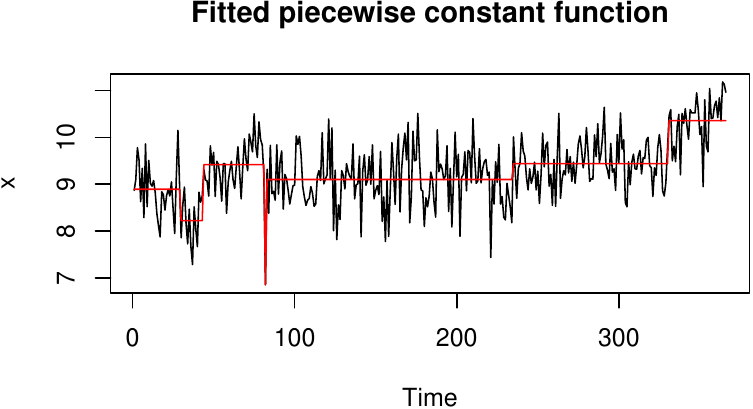}}

}

\subcaption{\label{fig-cet-old-2}WBS. Plot created by
\texttt{plot.wbs()}.}

\end{minipage}%
\newline
\begin{minipage}{0.50\linewidth}

\centering{

\pandocbounded{\includegraphics[keepaspectratio]{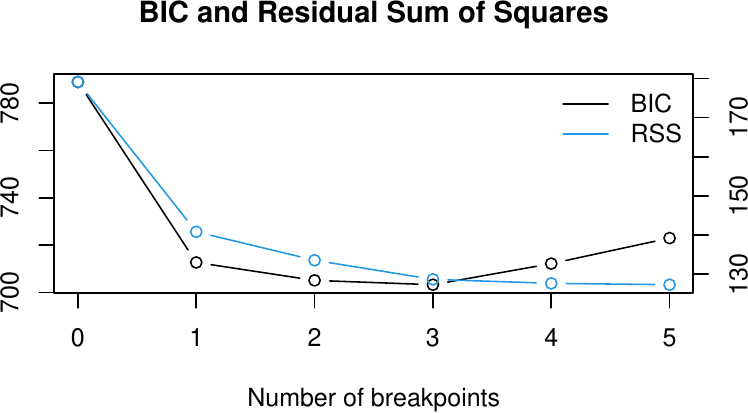}}

}

\subcaption{\label{fig-cet-old-3}strucchange. Plot created by
\texttt{plot.breakpointsfull()}.}

\end{minipage}%
\begin{minipage}{0.50\linewidth}

\centering{

\pandocbounded{\includegraphics[keepaspectratio]{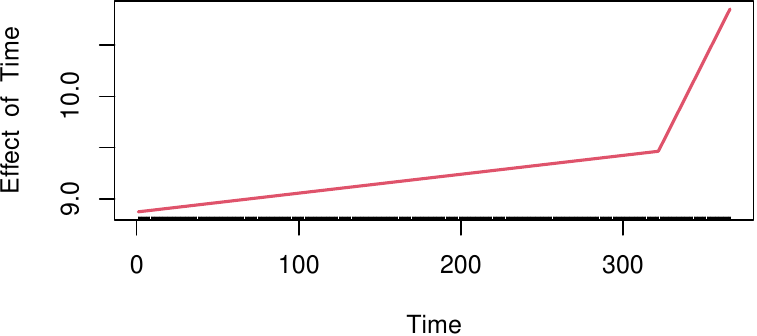}}

}

\subcaption{\label{fig-cet-old-4}segmented. Plot created by
\texttt{plot.segmented()}.}

\end{minipage}%

\caption{\label{fig-cet-old}Default plots returned by four different
changepoint detection packages. Note that some of the plots show the
original time series with changepoints overlaid, while others show
diagnostic plots.}

\end{figure}

Finally, you plot each object using \texttt{plot()}.
Figure~\ref{fig-cet-old} in our Appendix shows the resulting plots, some
of which add changepoints and model estimates to the original time
series, and others of which show diagnostic plots.

While you were able to conduct your analysis, you had to learn the
ins-and-outs of four different packages, all of which were designed in
their own way.

\subsection{An example using tidychangepoint}\label{sec-ex-tidycpt}

{tidychangepoint} makes the comparisons in the previous section simpler
by employing a unified syntax and output object. With {tidychangepoint},
it is possible to iterate analyses over lists of specifications.

We demonstrate this approach using the same four models as above, plus a
null model that finds no changepoints, and a genetic algorithm that Shi
et al. (2022) use to explore a variety of different parametric models
using genetic algorithms and a previous version of the CET data
set.\footnote{Note that we have set the \textbf{max}imum number of
  \textbf{iter}ations (generations) to just 1000. To replicate the exact
  results of Shi et al. (2022) we would need a much larger value. As
  noted in Section~\ref{sec-correctness}, the values in the underlying
  data set have also changed. However, we note that even with different
  data and many fewer generations, we identified all three of the
  changepoints found by Shi et al. (2022) to within a year (along with
  three others).} Using a trendshift model (see
Section~\ref{sec-models}), they identified the years 1700, 1739, and
1988 as changepoints.\footnote{Because this analysis leveraged the {GA}
  (Scrucca 2024) package, which is not specific to changepoint
  detection, each model that Shi et al. (2022) wanted to test required
  writing a different set of code, the complexity of which prevents us
  from reproducing it here.}.

\begin{Shaded}
\begin{Highlighting}[]
\NormalTok{cet\_objs }\OtherTok{\textless{}{-}} \FunctionTok{list}\NormalTok{(}
  \AttributeTok{null =} \FunctionTok{segment}\NormalTok{(CET, }\AttributeTok{method =} \StringTok{"null"}\NormalTok{),}
  \AttributeTok{pelt =} \FunctionTok{segment}\NormalTok{(CET, }\AttributeTok{method =} \StringTok{"pelt"}\NormalTok{),}
  \AttributeTok{wbs =} \FunctionTok{segment}\NormalTok{(CET, }\AttributeTok{method =} \StringTok{"wbs"}\NormalTok{),}
  \AttributeTok{bkp =} \FunctionTok{segment}\NormalTok{(CET, }\AttributeTok{method =} \StringTok{"strucchange"}\NormalTok{),}
  \AttributeTok{seg =} \FunctionTok{segment}\NormalTok{(CET, }\AttributeTok{method =} \StringTok{"segmented"}\NormalTok{),}
  \AttributeTok{shi =} \FunctionTok{segment}\NormalTok{(CET, }\AttributeTok{method =} \StringTok{"ga{-}shi"}\NormalTok{, }\AttributeTok{maxiter =} \DecValTok{1000}\NormalTok{, }\AttributeTok{run =} \DecValTok{100}\NormalTok{)}
\NormalTok{)}
\end{Highlighting}
\end{Shaded}

Once the algorithms have been run, we can easily extract the sets of
changepoints for comparison using the \texttt{changepoints()} function
and convert them to human-interpretable years with the
\texttt{as\_year()} function.

\begin{Shaded}
\begin{Highlighting}[]
\NormalTok{cet\_objs }\SpecialCharTok{|\textgreater{}}
  \FunctionTok{lapply}\NormalTok{(changepoints, }\AttributeTok{use\_labels =} \ConstantTok{TRUE}\NormalTok{) }\SpecialCharTok{|\textgreater{}}
  \FunctionTok{lapply}\NormalTok{(as\_year)}
\end{Highlighting}
\end{Shaded}

\begin{verbatim}
$null
character(0)

$pelt
[1] "1713" "1715" "1925" "2002" "2005"

$wbs
[1] "1701" "1739" "1740" "1892" "1988"

$bkp
[1] "1713" "1910" "1970"

$seg
[1] "1980"

$shi
[1] "1691" "1699" "1740" "1741" "1893" "1989"
\end{verbatim}

\begin{figure}[H]

\centering{

\pandocbounded{\includegraphics[keepaspectratio]{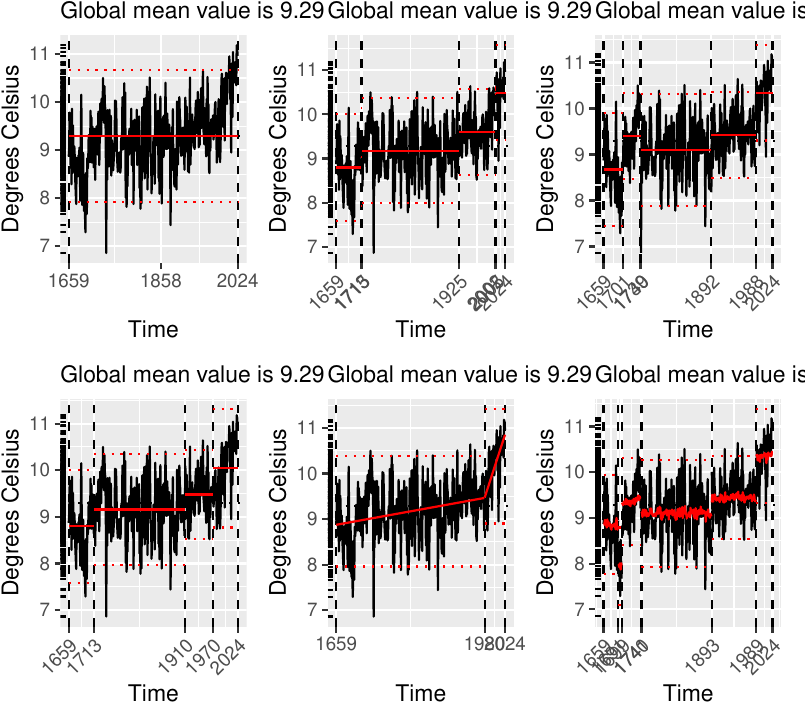}}

}

\caption{\label{fig-cet-compare}Comparison of four different attempts to
detect changepoints in the Central England temperature data. Note that
the model in the lower-right panel includes an autoregressive parameter,
and thus the model estimates vary across years within regions. All plots
created by \texttt{plot.tidycpt()}.}

\end{figure}

We compare the results visually using \texttt{plot()} (see
Figure~\ref{fig-cet-compare}), which now always situates the model
estimates in the domain of the original time series. Each panel of
Figure~\ref{fig-cet-compare} shows the original time series (black line)
segmented by the changepoints determined by the various algorithms
(vertical dotted lines). The horizontal red lines indicate the mean and
1.96 times the standard deviation within each region, as per the summary
produced by \texttt{tidy()} (see Section~\ref{sec-methods}).

\phantomsection\label{cell-fig-cet-compare}
\begin{Shaded}
\begin{Highlighting}[]
\NormalTok{cet\_objs }\SpecialCharTok{|\textgreater{}}
  \FunctionTok{lapply}\NormalTok{(plot, }\AttributeTok{use\_time\_index =} \ConstantTok{TRUE}\NormalTok{, }\AttributeTok{ylab =} \StringTok{"Degrees Celsius"}\NormalTok{) }\SpecialCharTok{|\textgreater{}}
\NormalTok{  patchwork}\SpecialCharTok{::}\FunctionTok{wrap\_plots}\NormalTok{()}
\end{Highlighting}
\end{Shaded}

We return to this example in Section~\ref{sec-cet}.

\subsection{Mathematical problem formulation}\label{sec-math}

Let \(y = \{y_1, \ldots, y_n\} \in \mathbb{R}^n\) be a discrete series
of observations made over an integer-valued index of time points
\(t = 1, \ldots, n\). Define
\(\tau = \{\tau_1, \ldots, \tau_m\} \in [1,n]^m\) to be a set of time
point indexes known as \emph{changepoints}, for some
\(0 \leq m \leq n\). The \(m\) changepoints divide the time series into
\(m+1\) regions, with the idea that the behavior of the time series
within each region does not change, and the changepoints represent
points in time at which the behavior of the underlying system changes in
some meaningful way. If we adopt the convention that \(\tau_0 = 1\) and
\(\tau_{m+1} = n+1\) then the union of the set of half-open intervals
\([\tau_j, \tau_{j+1})\) for \(0 \leq j \leq m\) covers the time domain
of the series \(y\).

Even if the changepoints are assumed to come from the discrete set
\(t\), there are \(2^n\) possible changepoint sets, and the Changepoint
Detection Problem (CDP)---which is the problem of finding the optimal
changepoint set (according to some metric)---has attracted considerable
attention in the literature. Aminikhanghahi and Cook (2017) provide an
excellent survey of relevant results. In the typical unsupervised
setting in which the true changepoints are not known, various models,
metrics, and algorithms to find the optimal changepoint set have been
proposed, some of which are in widespread use. We briefly highlight the
previous work most relevant to ours in Section~\ref{sec-algs}, and
review existing software implementations of these algorithms in
Section~\ref{sec-pkgs}. In any case, all attempts to solve the CDP
considered in this paper involve three essential components: 1) a
parametric model \(M(y | \theta)\) that describes the data-generating
process of the time series; 2) an algorithm \(A\) to search the
exponentially large space of possible changepoint sets; and 3) a metric
\(f : (\tau, M) \rightarrow \mathbb{R}\) (a penalized objective
function) for evaluating the suitability of any given changepoint set.
We define \(\tau_{f, M, \theta}^*\) to be the changepoint set that
minimizes the value of the penalized objective function \(f\) on the
model \(M\) with parameters \(\theta\) over the space of all possible
changepoint sets. That is, \[
  \tau_{f, M, \theta_\tau}^* = \arg\min_{\tau \in [1,n]^m} f(\tau, M(y | \theta_\tau)) 
\] The value of the penalized objective function
\(f(\tau, M(y|\hat{\theta}_\tau))\) is often referred to as the
\emph{fitness} of the changepoint set \(\tau\) and the model \(M\) with
estimated parameters \(\hat{\theta}_\tau\) to the time series \(y\).
Note that the value of the parameters \(\theta\) depend on the
changepoint set \(\tau\), so we may write \(\theta_\tau\).

For example, perhaps the simplest and most common parametric model for
changepoint detection is the normal (Gaussian) \emph{meanshift} model,
in which the time series \(y\) is assumed to be generated by a different
random variable \(Y_j \sim \mathcal{N}(\mu_j, \sigma^2)\) for each
region \(1 \leq j \leq m+1\). Thus the series mean \(\mu_j\) is constant
within each of the \(m+1\) regions, but varies across the regions
(piecewise stationarity), while the variance \(\sigma^2\) is constant
over the entire time series. We fit the normal meanshift model using
Maximium Likelihood Estimation (MLE), resulting in the estimated
parameters
\(\hat{\theta} = \{ \hat{\sigma}^2, \hat{\mu}_0, \ldots, \hat{\mu}_m \}\).
If we choose the Bayesian Information Criterion (BIC) (Schwarz 1978) as
our penalty, then the penalized objective function becomes: \[
  BIC(\tau, M(y|\hat{\theta}_\tau)) = k_M(\tau) \ln{n} - 2 \ln{L_M(y | \hat{\theta}_\tau)},
\] where \(k_M(\tau)\) is the number of parameters estimated by the
model (in this case, \(m + 2\)) and \(L_M(y|\hat{\theta}_\tau)\) is the
likelihood function for \(M\), which in this case is (see S. Li and Lund
(2012) for details): \[
  L_M(y|\hat{\theta}_\tau) = -\frac{n}{2}(\ln{\hat{\sigma}^2} + 1 + \ln{2 \pi}) \,.
\]

In general, while the true value of
\(\tau_{BIC, meanshift, \hat{\theta}}^*\) is unknown, under reasonable
assumptions about the growth of the penalized objective function with
respect to the number of changepoints, the PELT algorithm (Killick,
Fearnhead, and Eckley 2012) will find the exact value in polynomial
time.

More generally, given a time series \(y\), a parametric model
\(M(y|\theta)\), and a penalized objective function \(f\), the goal of
any changepoint detection algorithm \(A\) is to search the exponentially
large space of possible changepoints sets for the one \(\tau^*\) that
minimizes the value of \(f(\tau, M(y|\hat{\theta}_\tau))\). Of course,
in the process it must also estimate \(\hat{\theta}_\tau\).

\subsection{Selected changepoint detection algorithms}\label{sec-algs}

Aminikhanghahi and Cook (2017) provide a comprehensive survey of
different models for changepoint detection in time series. This includes
classifications of offline vs.~online algorithms, supervised
vs.~unsupervised problems, computational concerns, robustness, open
problems, and the distribution of changepoints within long time series.
Guédon (2015) quantifies the uncertainty in the segmentation of diverse
sets of changepoints, using both Bayesian and non-Bayesian techniques.

The Pruned Exact Linear Time (PELT) algorithm developed by Killick,
Fearnhead, and Eckley (2012) builds on a dynamic programming algorithm
from Jackson et al. (2005) to offer a linear time, exact algorithm for
changepoint detection. PELT offers order of magnitude improvements in
computational cost (i.e., \(O(n)\) vs.~the \(O(n \log{n})\) performance
of the binary segmentation algorithm (Scott and Knott 1974) and the
\(O(n^3)\) performance of the segmented neighborhood algorithm (Auger
and Lawrence 1989)), and is capable of optimizing over different
parametric models (e.g., meanvar, meanshift, etc.) and different
penalized objective functions (e.g., BIC, MBIC (see
Section~\ref{sec-mbic}), etc.). To guarantee an optimal solution, PELT
requires that the value of the penalized objective function decreases
(or increases by an amount bounded by a known constant) with each
additional changepoint. Moreover, its running time can be quadratic
(i.e., \(O(n^2)\)) unless the number of changepoints grows linearly with
the length of the time series. While these assumptions are mild, not all
possible combinations of time series, models, and penalty functions will
meet them.\footnote{To see why, note that PELT's reliance on dynamic
  programming requires that the penalized objective function be additive
  across the regions. Some penalty functions, such as MDL (see
  Section~\ref{sec-mdl}) depend on the \emph{locations} of the
  changepoints, and thus cannot be additive across the regions.}

S. Li and Lund (2012) illustrate how a genetic algorithm can provide
good results by employing Darwinian principles (e.g., survival of the
fittest) to search the space of all possible changepoint sets
intelligently. For a given parametric model and penalized objective
function, a genetic algorithm starts with an initial ``generation'' of
(usually randomly generated) candidate changepoint sets, evaluates them
all using the penalized objective function, then probabilistically
``mates'' the most promising candidate changepoint sets to produce a new
generation of the same size. This process is repeated until a stopping
criteria is met, and the single best (fittest) changepoint set is
returned. While genetic algorithms are neither exact, nor deterministic,
nor computationally efficient, they are more easily generalizable than
PELT, and can return competitive results in practice (Shi et al. 2022).
S. Li and Lund (2012) also develop the notion of the Minimum Descriptive
Length (MDL) as a penalty function based on information theory.

Bai and Perron (1998) use linear regression models fit by least squares
to detect changepoints. In particular, they construct a framework for
testing the alternative hypothesis that there is one additional
changepoint. Zeileis et al. (2003) build on this and related work by Bai
and Perron (2003), using a series of F-statistics and a dynamic
programming algorithm to find the set of changepoints that minimize the
residual sum of squared errors (see Section~\ref{sec-pkgs}). V. M.
Muggeo (2003) also employs a linear regression framework, but
iteratively fit models using the Davies test (Davies 1987) as a
criterion.

Barry and Hartigan (1993) employ Bayesian techniques for changepoint
detection. Cho and Fryzlewicz (2015) propose the Sparsified Binary
Segmentation (SBS) algorithm, which uses CUSUM statistics to analyze
high dimensional time series. This algorithm uses the concept of a
threshold above which statistics are kept---an idea that resurfaces in
Suárez-Sierra, Coen, and Taimal (2023). Cho (2016) explores the utility
of a double CUSUM statistic for panel data. Hocking et al. (2013) use
machine learning techniques to determine the optimal number of
changepoints.

Suárez-Sierra, Coen, and Taimal (2023) detail the implementation of a
changepoint detection heuristic we rebrand as Coen's algorithm in this
paper. Coen's algorithm (Section~\ref{sec-coen}) is genetic and uses a
non-homogeneous Poisson process model (Section~\ref{sec-nhpp}) to model
the exceedances of a threshold over time, and a Bayesian Minimum
Descriptive Length (BMDL, Section~\ref{sec-bmdl}) penalized objective
function. Taimal, Suárez-Sierra, and Rivera (2023) discuss modifications
and performance characteristics of Coen's algorithm.

\subsection{Existing changepoint detection R packages}\label{sec-pkgs}

There are dozens of changepoint packages on CRAN, all of which have
their strengths and limitations. Please see
\href{https://lindeloev.github.io/mcp/articles/packages.html}{the
``Comparison to other packages'' vignette} in the {mcp} (Lindeløv 2024)
documentation for a thorough comparison.

According to statistics returned by the {pkgsearch} (Csárdi and Salmon
2025) package, only three changepoint detection packages have more than
1000 downloads per month and more than two reverse dependencies:
{strucchange} (Zeileis et al. 2024), {segmented} (V. M. R. Muggeo 2025),
and {changepoint} (Killick 2024; Killick and Eckley 2014).
{tidychangepoint} now supports all three.

{strucchange} (Zeileis et al. 2024, 2002) uses a regression-based
framework and implements a dynamic programming algorithm to identify
changepoints. The {segmented} (V. M. R. Muggeo 2025) package iteratively
searches for optimal changepoint sets in a linear regression framework.

{changepoint} (Killick 2024; Killick and Eckley 2014) implements the
algorithms PELT, Binary Segmentation, and Segmented Neighborhood. PELT
employs one of three different models: meanshift, meanvar, and varshift
(see Section~\ref{sec-models}). For the meanvar model (the most
versatile) the data-generating model can follow a Normal, Poisson,
Gamma, or Exponential distribution. Penalty functions (see
Section~\ref{sec-penalty}) implemented by PELT include AIC, BIC, and
MBIC, but not MDL. {EnvCpt} (Killick et al. 2025) is another package
that wraps {changepoint} functionality.

The {wbs} (Baranowski and Fryzlewicz 2024) package implements the Wild
Binary Segmentation and standard Binary Segmentation algorithms, the
first of which is incorporated into {tidychangepoint}. The
{ggchangepoint} (Yu 2022) package provides some wrapper functions and
graphical tools for changepoint detection analysis that are similar to
ours, but more limited in scope and usability. Nonparametric methods for
changepoint detection are present in the {ecp} (James, Zhang, and
Matteson 2024) and {changepoint.np} (Haynes and Killick 2022) packages.
We consider these beyond our scope at this time, which is focused on
parameteric models for changepoint detection.

A Bayesian approach is implemented in the {bcp} (X. Wang, Erdman, and
Emerson 2018; Erdman and Emerson 2008) package, which is no longer
available on CRAN. The {qcc} (Scrucca 2017) package is more broadly
focused on quality control and statistical process control, neither of
which are specific to changepoint detection. The {mcp} (Lindeløv 2024)
package focuses on regression-based, Bayesian methods.

The {changepointGA} (M. Li and Lu 2025) package appeared on CRAN
subsequent to the submission of this paper. It provides a detailed
implementation of several changepoint detection algorithms that are
built on top of a custom algorithmic implementation (M. Li and Lu 2024).
Note that in contrast, {tidychangepoint} relies on a more
general-purpose genetic algorithm from the {GA} (Scrucca 2024) package
as a backend. We have included basic support for the ARIMA model with a
BIC penalty function in {tidychangepoint}.

\subsection{Our contribution}\label{our-contribution}

In this paper, we present several substantive improvements to the
existing state-of-the-art. First, the {tidychangepoint} package (Baumer
et al. 2025) provides a unified interface that makes working with
existing changepoint detection algorithms in a {tidyverse}-compliant
syntax easy (Section~\ref{sec-tidychangepoint}). This reduces friction
for anyone who wants to compare the results of algorithms, models, or
penalized objective functions across different packages. Second, the
architecture of the package is easily extended to incorporate new
changepoint detection algorithms (Section~\ref{sec-alg}), new parametric
models (Section~\ref{sec-models}), and new penalized objective functions
(Section~\ref{sec-penalty}). Indeed, our careful separation of these
three essential elements of changepoint detection is in and of itself
clarifying. Third, to the best of our knowledge, {tidychangepoint} is
the first R package that allows users to run genetic algorithms for
changepoint detection without having to write custom code. With {GA} as
a backend, {tidychangepoint} provides a user-friendly interface for
mixing-and-matching various parametric models, penalized objective
functions, and initial populations in ways that previously required
non-trivial programming by the user. Since {tidychangepoint} wraps
functionality from other packages it adds only modest computational
overhead, and delivers correct results (Section~\ref{sec-results}).
Finally, {tidychangepoint} includes new or updated data sets packaged as
time series objects about particulate matter in Bogotá, monthly rainfall
in Medellín, temperature in Central England, and batting performance in
Major League Baseball. We illustrate the utility of the package in
Section~\ref{sec-examples} and conclude in Section~\ref{sec-conclusion}.

{tidychangepoint} is available via download from the Comprehensive R
Archive Network (CRAN) since August 19, 2024. The package website
contains several short articles demonstrating how to use various
features of the package, some of which are not included in this paper
(for brevity).

\section{Examples using tidychangepoint}\label{sec-examples}

In this Section we provide more detailed examples of how
{tidychangepoint} can facilitate meaningful comparisons across
algorithms, models, and penalized objective functions.

\subsection{Temperatures in Central England}\label{sec-cet}

In Section~\ref{sec-ex-tidycpt}, we noted that the comparative analysis
across different models presented in Shi et al. (2022) required writing
custom code for each model fit. In Section~\ref{sec-models}, we list the
many parametric models that {tidychangepoint} includes. These include
the ``meanshift'' (i.e., piecewise-constant) and ``trendshift'' (i.e.,
piecewise-linear) models, either of which can include AR(1) lagged
errors, and either of which can be fit using the BIC or MDL penalties
(see Section~\ref{sec-penalty}).

With {tidychangepoint}, we can employ a machine learning style approach
to finding the best combination of model and penalty function for a time
series. In this case, we build a \texttt{data.frame} with 8 possible
combinations.

\begin{Shaded}
\begin{Highlighting}[]
\NormalTok{cet\_options }\OtherTok{\textless{}{-}} \FunctionTok{expand\_grid}\NormalTok{(}
  \AttributeTok{model\_fn =} \FunctionTok{c}\NormalTok{(fit\_meanshift\_norm, fit\_meanshift\_norm\_ar1, }
\NormalTok{               fit\_trendshift, fit\_trendshift\_ar1),}
  \AttributeTok{penalty\_fn =} \FunctionTok{c}\NormalTok{(BIC, MDL)}
\NormalTok{)}
\end{Highlighting}
\end{Shaded}

Next, using functionality\footnote{\texttt{purrr::map2()} is similar to
  \texttt{base::mapply()}.} from the {purrr} package (Wickham and Henry
2025), we simply iterate the \texttt{segment()} function---which is the
main workhorse in {tidychangepoint}---over these combinations. In each
case we employ a genetic algorithm that can produce up to 5000
generations (\texttt{maxiter}), although it will stop if there is no
improvement in the fitness (i.e., the value of the penalized objective
function) after 100 consecutive generations (\texttt{run}).

\begin{Shaded}
\begin{Highlighting}[]
\FunctionTok{library}\NormalTok{(purrr)}
\NormalTok{cet\_mods }\OtherTok{\textless{}{-}}\NormalTok{ cet\_options }\SpecialCharTok{|\textgreater{}}
  \FunctionTok{mutate}\NormalTok{(}
    \AttributeTok{tidycpt =} \FunctionTok{map2}\NormalTok{(}
\NormalTok{      model\_fn, penalty\_fn, }\AttributeTok{.f =}\NormalTok{ segment, }
      \AttributeTok{x =}\NormalTok{ CET, }\AttributeTok{method =} \StringTok{"ga"}\NormalTok{, }\AttributeTok{maxiter =} \DecValTok{5000}\NormalTok{, }\AttributeTok{run =} \DecValTok{100}
\NormalTok{    )}
\NormalTok{  )}
\end{Highlighting}
\end{Shaded}

The \texttt{fitness()}, \texttt{changepoints()} and
\texttt{model\_name()} functions can recover useful information from the
resulting objects (see Section~\ref{sec-methods}).

\begin{Shaded}
\begin{Highlighting}[]
\NormalTok{cet\_mods }\OtherTok{\textless{}{-}}\NormalTok{ cet\_mods }\SpecialCharTok{|\textgreater{}}
  \FunctionTok{mutate}\NormalTok{(}
    \AttributeTok{fitness =} \FunctionTok{map\_dbl}\NormalTok{(tidycpt, fitness),}
    \AttributeTok{penalty =} \FunctionTok{map\_chr}\NormalTok{(tidycpt, }\SpecialCharTok{\textasciitilde{}}\FunctionTok{names}\NormalTok{(}\FunctionTok{fitness}\NormalTok{(.x))),}
    \AttributeTok{num\_cpts =} \FunctionTok{map\_int}\NormalTok{(tidycpt, }\SpecialCharTok{\textasciitilde{}}\FunctionTok{length}\NormalTok{(}\FunctionTok{changepoints}\NormalTok{(.x))),}
    \AttributeTok{model\_name =} \FunctionTok{map\_chr}\NormalTok{(tidycpt, model\_name),,}
    \AttributeTok{num\_gens =} \FunctionTok{map\_int}\NormalTok{(tidycpt, }\SpecialCharTok{\textasciitilde{}}\FunctionTok{nrow}\NormalTok{(}\FunctionTok{as.segmenter}\NormalTok{(.x)}\SpecialCharTok{@}\NormalTok{summary))}
\NormalTok{  )}
\end{Highlighting}
\end{Shaded}

We can now easily compare the results and learn about which model
performed best with which penalty function.

\begin{Shaded}
\begin{Highlighting}[]
\NormalTok{cet\_mods }\SpecialCharTok{|\textgreater{}}
  \FunctionTok{arrange}\NormalTok{(penalty, fitness) }\SpecialCharTok{|\textgreater{}}
  \FunctionTok{select}\NormalTok{(model\_name, fitness, penalty, num\_cpts, num\_gens)}
\end{Highlighting}
\end{Shaded}

\begin{verbatim}
# A tibble: 8 x 5
  model_name         fitness penalty num_cpts num_gens
  <chr>                <dbl> <chr>      <int>    <int>
1 trendshift            671. BIC            3     1246
2 meanshift_norm_ar1    674. BIC            6     1362
3 meanshift_norm        680. BIC            9     1415
4 trendshift_ar1        682. BIC            4     1347
5 trendshift_ar1        662. MDL            5     2987
6 meanshift_norm        680. MDL            8     1480
7 meanshift_norm_ar1    683. MDL            5     1451
8 trendshift            702. MDL            9     1754
\end{verbatim}

In this case, the trendshift model with white noise errors produces the
best fitness score with the BIC penalty, while adding AR(1) errors
produces the best fitness score with the MDL penalty. These results
accord with the findings by Shi et al. (2022).

\subsection{The designated hitter in Major League
Baseball}\label{sec-mlb}

\begin{figure}[H]

\centering{

\pandocbounded{\includegraphics[keepaspectratio]{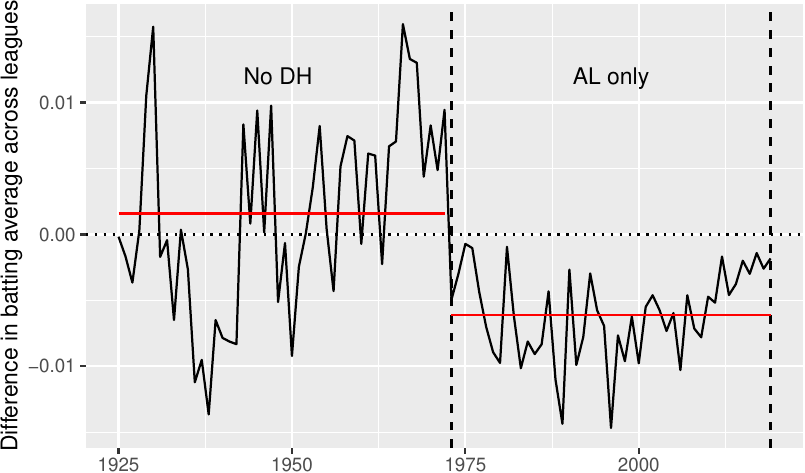}}

}

\caption{\label{fig-mlb-hrs}Difference in the rate of home runs per
plate appearance between the American and National Leagues in Major
League Baseball, 1925--2019. The designated hitter rule was adopted in
1973 by the American League, but not until 2020 by the National League.
Note how, during the period from 1973 to 2019 when only the American
League employed a designated hitter, the difference in batting average
was always negative.}

\end{figure}

Sometimes we know the truth about when changepoints occur. For example,
Figure~\ref{fig-mlb-hrs} shows the difference in batting average between
the American and National Leagues of Major League Baseball (MLB) from
1925--2019. The introduction of the designated hitter rule in the
American League in 1973 led to noticeable differences in batting
performance relative to the National League (which did not adapt the
rule until 2020), because the rule allowed teams in the American League
to designate another player to hit in place of the pitcher, whereas in
the National League, the current pitcher continued to be forced to hit.
Most pitchers are poor hitters relative to a Major League non-pitcher,
and so the designated hitter rule allowed American League teams to bat 9
non-pitchers all the time, whereas National League teams generally had 8
non-pitchers and one pitcher in the batting lineup. The goal of the rule
change was to increase offensive performance in the American League, and
it worked. Note that in Figure~\ref{fig-mlb-hrs}, the American League
had a higher batting average in every single year from 1973 to 2019.
Thus, the year 1973 is a known changepoint, because the data generating
process changed before and after that year.

\phantomsection\label{cell-fig-mlb-hrs}

Using the \texttt{segment()} function, we create a list of
\texttt{tidycpt} model objects (see Section~\ref{sec-tidychangepoint}).
We apply the PELT algorithm for a change in means and variances using
the default MBIC penalty function, and also try a model that allows only
the mean to vary across regions and uses the BIC penalty function. Next,
we try the WBS algorithm. Finally, we try a genetic algorithm using a
normal meanshift model and the MDL penalty function (see
Section~\ref{sec-penalty}). This algorithm will stop after 500
generations (\texttt{maxiter}), or after 50 consecutive generations with
no improvement (\texttt{run}), whichever comes first.

\begin{Shaded}
\begin{Highlighting}[]
\NormalTok{mlb\_mods }\OtherTok{\textless{}{-}} \FunctionTok{list}\NormalTok{(}
  \StringTok{"pelt\_meanvar"} \OtherTok{=} \FunctionTok{segment}\NormalTok{(mlb\_bavg, }\AttributeTok{method =} \StringTok{"pelt"}\NormalTok{),}
  \StringTok{"pelt\_meanshift"} \OtherTok{=} \FunctionTok{segment}\NormalTok{(}
\NormalTok{    mlb\_bavg, }\AttributeTok{method =} \StringTok{"pelt"}\NormalTok{, }
    \AttributeTok{model\_fn =}\NormalTok{ fit\_meanshift\_norm, }\AttributeTok{penalty =} \StringTok{"BIC"}
\NormalTok{  ),}
  \StringTok{"wbs"} \OtherTok{=} \FunctionTok{segment}\NormalTok{(mlb\_bavg, }\AttributeTok{method =} \StringTok{"wbs"}\NormalTok{),}
  \StringTok{"ga"} \OtherTok{=} \FunctionTok{segment}\NormalTok{(}
\NormalTok{    mlb\_bavg, }\AttributeTok{method =} \StringTok{"ga"}\NormalTok{, }\AttributeTok{model\_fn =}\NormalTok{ fit\_meanshift\_norm, }
    \AttributeTok{penalty\_fn =}\NormalTok{ MDL, }\AttributeTok{maxiter =} \DecValTok{500}\NormalTok{, }\AttributeTok{run =} \DecValTok{50}
\NormalTok{  )}
\NormalTok{)}
\end{Highlighting}
\end{Shaded}

Because each of the resulting objects are of class \texttt{tidycpt} (see
Section~\ref{sec-tidychangepoint}), comparing the results across the
four algorithms is easy using \texttt{map()}\footnote{\texttt{purrr::map()}
  is similar to \texttt{base::lapply()}.} (from the {purrr} package
(Wickham and Henry 2025)). First, we examine the changepoint sets.

\begin{Shaded}
\begin{Highlighting}[]
\NormalTok{mlb\_mods }\SpecialCharTok{|\textgreater{}}
  \FunctionTok{map}\NormalTok{(changepoints, }\AttributeTok{use\_labels =} \ConstantTok{TRUE}\NormalTok{) }\SpecialCharTok{|\textgreater{}}
  \FunctionTok{map}\NormalTok{(as\_year)}
\end{Highlighting}
\end{Shaded}

\begin{verbatim}
$pelt_meanvar
[1] "1976" "2011"

$pelt_meanshift
character(0)

$wbs
[1] "1928" "1930" "1942" "1956" "1972"

$ga
 [1] "1929" "1931" "1936" "1943" "1948" "1952" "1966" "1969" "1973" "1978"
[11] "2012"
\end{verbatim}

PELT finds two changepoints, including 1976, which is just three years
after the true changepoint of 1973. It also finds 2011 to be a
changepoint, which is interesting because the difference in batting
average climbs noticeably around this time. PELT using the meanshift
model and the BIC penalty finds no changepoints. WBS finds 5
changepoints, including 1972, just one year before the true changepoint.
Our genetic algorithm finds 11 changepoints, including the true
changepoint.

To compare other aspects of the fits, we can iterate the
\texttt{glance()} function from the {broom} package (Robinson et al.
2025) over that same list. Since \texttt{glance()} always returns a
one-row \texttt{tibble}, we can then combine those \texttt{tibble}s into
one using the \texttt{list\_rbind()} function. Care must be taken to
ensure that these comparisons are meaningful. While comparing, say, the
elapsed time across these algorithms and models is fair, comparing the
fitness values is more fraught. MDL scores are not comparable to BIC or
MBIC scores, for example.

\begin{Shaded}
\begin{Highlighting}[]
\NormalTok{mlb\_mods }\SpecialCharTok{|\textgreater{}}
  \FunctionTok{map}\NormalTok{(glance) }\SpecialCharTok{|\textgreater{}}
  \FunctionTok{list\_rbind}\NormalTok{()}
\end{Highlighting}
\end{Shaded}

\begin{verbatim}
# A tibble: 4 x 8
  pkg      version algorithm seg_params model_name criteria fitness elapsed_time
  <chr>    <pckg_> <chr>     <list>     <chr>      <chr>      <dbl> <drtn>      
1 changep~ 2.3     PELT      <list [1]> meanvar    MBIC       -745.  0.011 secs 
2 changep~ 2.3     PELT      <list [1]> meanshift~ BIC        -668.  0.002 secs 
3 wbs      1.4.1   Wild Bin~ <list [1]> meanshift~ MBIC       -473.  0.010 secs 
4 GA       3.2.4   Genetic   <list [1]> meanshift~ MDL        -735. 19.982 secs 
\end{verbatim}

\begin{figure}

\begin{minipage}{0.50\linewidth}

\centering{

\pandocbounded{\includegraphics[keepaspectratio]{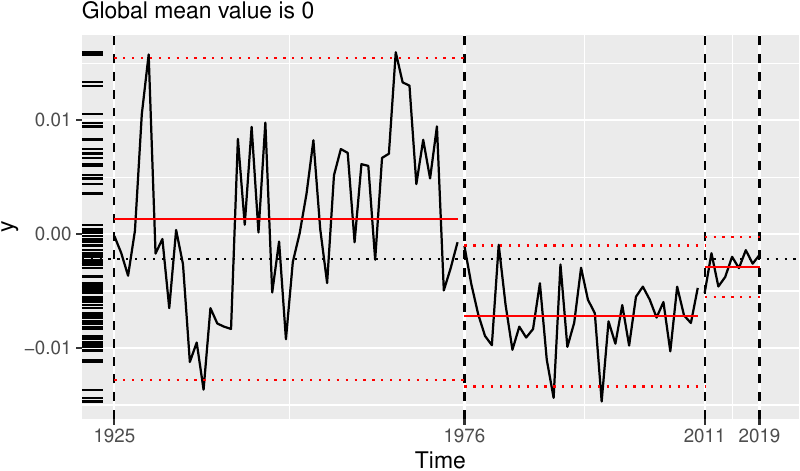}}

}

\subcaption{\label{fig-mlb-ga-1}The PELT algorithm using the mean and
variance model and the MBIC penalty.}

\end{minipage}%
\begin{minipage}{0.50\linewidth}

\centering{

\pandocbounded{\includegraphics[keepaspectratio]{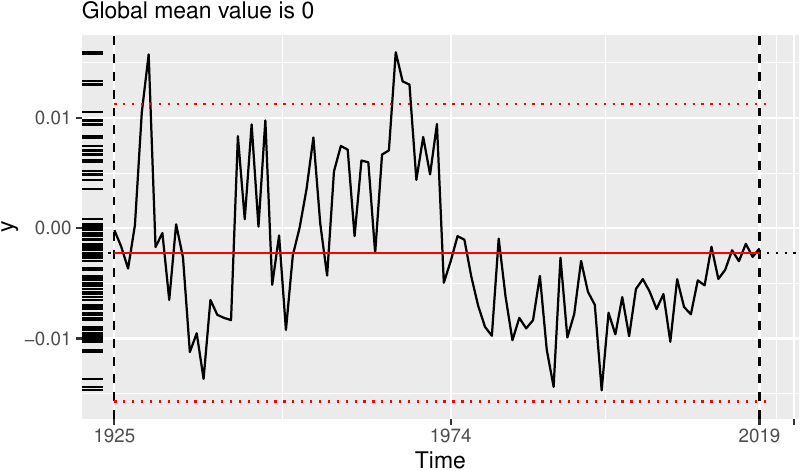}}

}

\subcaption{\label{fig-mlb-ga-2}The PELT algorithm using the meanshift
model and the BIC penalty.}

\end{minipage}%
\newline
\begin{minipage}{0.50\linewidth}

\centering{

\pandocbounded{\includegraphics[keepaspectratio]{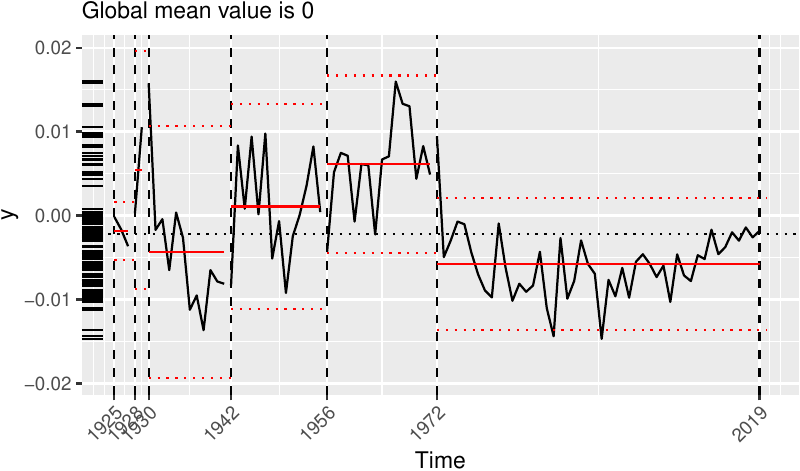}}

}

\subcaption{\label{fig-mlb-ga-3}The WBS algorithm.}

\end{minipage}%
\begin{minipage}{0.50\linewidth}

\centering{

\pandocbounded{\includegraphics[keepaspectratio]{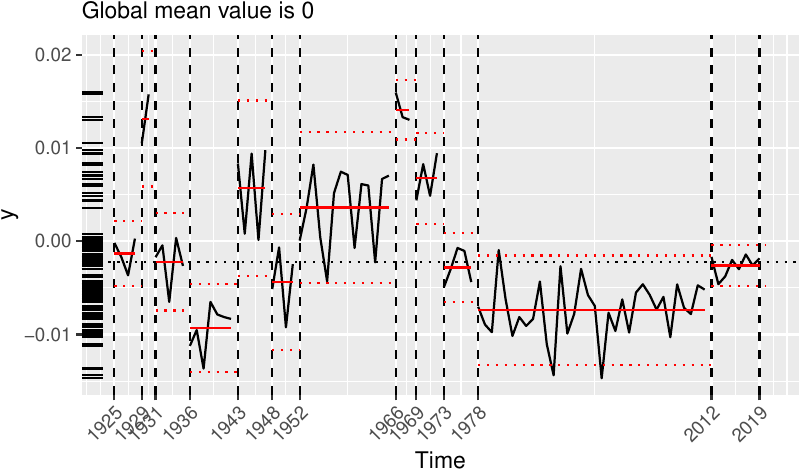}}

}

\subcaption{\label{fig-mlb-ga-4}A genetic algorithm using the meanshift
model and the MDL penalty.}

\end{minipage}%

\caption{\label{fig-mlb-ga}Comparison of changepoint sets returned by
four algorithms. The question of which provides the best fit is a matter
of interpretation.}

\end{figure}

In Figure~\ref{fig-mlb-ga}, we compare the changepoint sets found by the
four algorithms. While several of the algorithms indicate a changepoint
close to the true changepoint of 1973, it is debatable which provides
the best fit to the data.

\begin{figure}[H]

\centering{

\pandocbounded{\includegraphics[keepaspectratio]{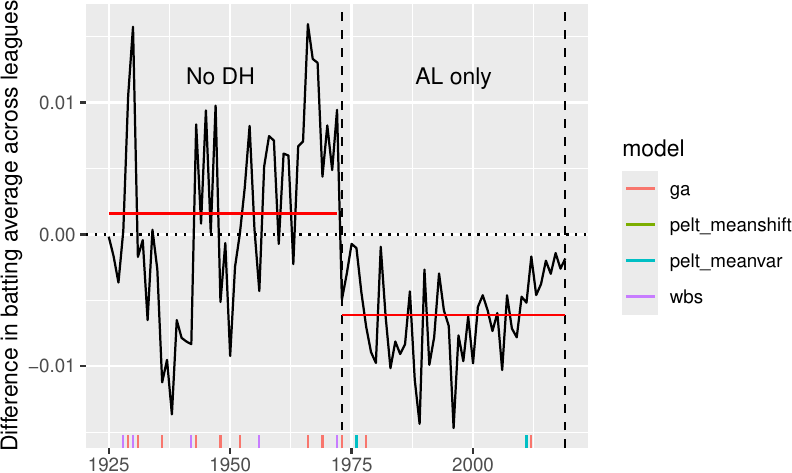}}

}

\caption{\label{fig-rug}An update to Figure~\ref{fig-mlb-hrs} adding a
rug for the distribution of changepoints found by the four algorithms
mentioned in this section.}

\end{figure}

Visually, we could use a rug plot (see \texttt{ggplot2::geom\_rug()}) to
show the distribution of changepoints across algorithms (see
Figure~\ref{fig-rug}). Three of the four algorithms found a changepoint
near the true changepoint of 1973. The other changepoints are false
positives with respect to the designated hitter rule, but may very well
reflect other changes in batting performance that we simply haven't been
thinking about. For example, we hypothesize that the recent decrease in
the difference in batting average across the two leagues could be a
result of the increased use of relief pitchers (who are almost always
replaced by a pinch hitter when their turn at bat comes up) at the
expense of starting pitchers (who can't be pinch-hit for because they
have to stay in the game)---a long-term change in \emph{pitching}
strategy. If this trend continued to its extreme and pitchers
\emph{never} batted, the designated hitter rule would be moot and the
American League's advantage would evaporate.

\phantomsection\label{cell-fig-rug}

\subsubsection{Investigating algorithmic
performance}\label{investigating-algorithmic-performance}

Because {tidychangepoint} provides model-fitting functions and penalized
objective functions that ``just work'', it is easy to investigate why
candidate changepoint sets of interest may not have been chosen by the
algorithm.

Because PELT is known to be optimal under mild conditions, all other
candidate changepoint sets should return a value for MBIC that is worse
than the one found by PELT (which includes the two changepoints 1976 and
2011). The \texttt{fitness()} function gives us this value.

\begin{Shaded}
\begin{Highlighting}[]
\FunctionTok{fitness}\NormalTok{(mlb\_mods[[}\StringTok{"pelt\_meanvar"}\NormalTok{]])}
\end{Highlighting}
\end{Shaded}

\begin{verbatim}
     MBIC 
-744.8353 
\end{verbatim}

As expected, the MBIC value of the meanvar model with the true
changepoint set is worse (further from \(-\infty\)).

\begin{Shaded}
\begin{Highlighting}[]
\FunctionTok{MBIC}\NormalTok{(}\FunctionTok{fit\_meanvar}\NormalTok{(mlb\_bavg, }\AttributeTok{tau =} \DecValTok{49}\NormalTok{))}
\end{Highlighting}
\end{Shaded}

\begin{verbatim}
[1] -700.8384
\end{verbatim}

Why didn't PELT pick the changepoint set returned by the genetic
algorithm? This set, too, had a worse MBIC score, although its score is
better than the true changepoint set.

\begin{Shaded}
\begin{Highlighting}[]
\FunctionTok{MBIC}\NormalTok{(}\FunctionTok{fit\_meanvar}\NormalTok{(mlb\_bavg, }\AttributeTok{tau =} \FunctionTok{changepoints}\NormalTok{(mlb\_mods[[}\StringTok{"ga"}\NormalTok{]])))}
\end{Highlighting}
\end{Shaded}

\begin{verbatim}
[1] -725.6102
\end{verbatim}

Alternatively, we might ask why the genetic algorithm didn't pick either
the true changepoint set or the one found by PELT. Recall that the
genetic algorithm is randomized and only a heuristic, and so unlike in
the case of PELT, it is possible that we could manually identify a
changepoint set with a lower MDL than the one found by the algorithm.
However, in this case, the MDL score for the changepoint set returned by
the genetic algorithms bests the other two possibilities.

\begin{Shaded}
\begin{Highlighting}[]
\FunctionTok{fitness}\NormalTok{(mlb\_mods[[}\StringTok{"ga"}\NormalTok{]])}
\end{Highlighting}
\end{Shaded}

\begin{verbatim}
      MDL 
-735.2082 
\end{verbatim}

\begin{Shaded}
\begin{Highlighting}[]
\FunctionTok{list}\NormalTok{(}\DecValTok{49}\NormalTok{, }\FunctionTok{changepoints}\NormalTok{(mlb\_mods[[}\StringTok{"pelt\_meanvar"}\NormalTok{]])) }\SpecialCharTok{|\textgreater{}}
  \FunctionTok{map}\NormalTok{(fit\_meanshift\_norm, }\AttributeTok{x =}\NormalTok{ mlb\_bavg) }\SpecialCharTok{|\textgreater{}}
  \FunctionTok{map\_dbl}\NormalTok{(MDL)}
\end{Highlighting}
\end{Shaded}

\begin{verbatim}
[1] -691.7307 -682.8757
\end{verbatim}

This type of manual investigation is very difficult to do with other
changepoint detection packages.

\subsection{Particular matter in Bogotá, Colombia}\label{sec-bogota}

\begin{figure}[H]

\centering{

\pandocbounded{\includegraphics[keepaspectratio]{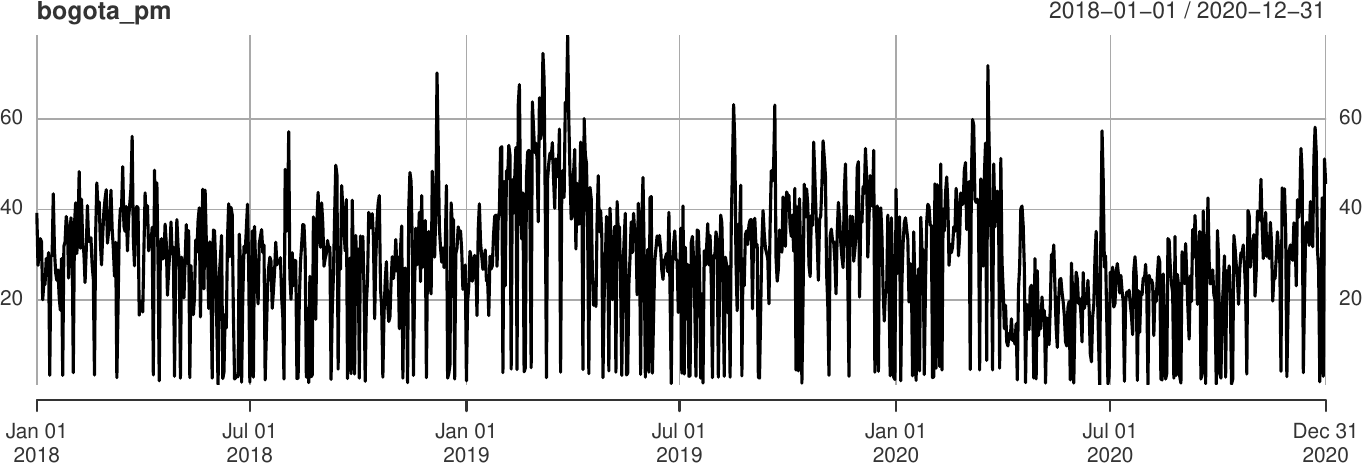}}

}

\caption{\label{fig-bogota}Particular matter above 2.5 microns in
Bogotá, Colombia, recorded daily from 2018--2020.}

\end{figure}

Consider the time series on particulate matter in Bogotá, Colombia
collected daily from 2018--2020, discussed in Suárez-Sierra, Coen, and
Taimal (2023), and shown in Figure~\ref{fig-bogota}. These data are
included in {tidychangepoint} as \texttt{bogota\_pm}.

\phantomsection\label{cell-fig-bogota}

There appears to be some seasonality in the particulate matter data as
well as a pronounced drop in particulate matter around the time of the
COVID-19 lockdown in March of 2020. Suárez-Sierra, Coen, and Taimal
(2023) use Coen's algorithm (see Section~\ref{sec-coen}) with a
manually-specified threshold of 37 \(\mu g / m^3\) informed by Colombian
government policy. Here, we reproduce that result using the {GA}-based
implementation in {tidychangepoint}. Although we find a different
changepoint set, the fit of the NHPP model is of comparable quality.

\begin{Shaded}
\begin{Highlighting}[]
\NormalTok{bog\_coen }\OtherTok{\textless{}{-}}\NormalTok{ bogota\_pm }\SpecialCharTok{|\textgreater{}}
  \FunctionTok{segment}\NormalTok{(}
    \AttributeTok{method =} \StringTok{"ga{-}coen"}\NormalTok{, }\AttributeTok{maxiter =} \DecValTok{1000}\NormalTok{, }\AttributeTok{run =} \DecValTok{100}\NormalTok{, }
    \AttributeTok{model\_fn\_args =} \FunctionTok{list}\NormalTok{(}\AttributeTok{threshold =} \DecValTok{37}\NormalTok{)}
\NormalTok{  )}
\end{Highlighting}
\end{Shaded}

\begin{figure}

\begin{minipage}{0.50\linewidth}

\centering{

\pandocbounded{\includegraphics[keepaspectratio]{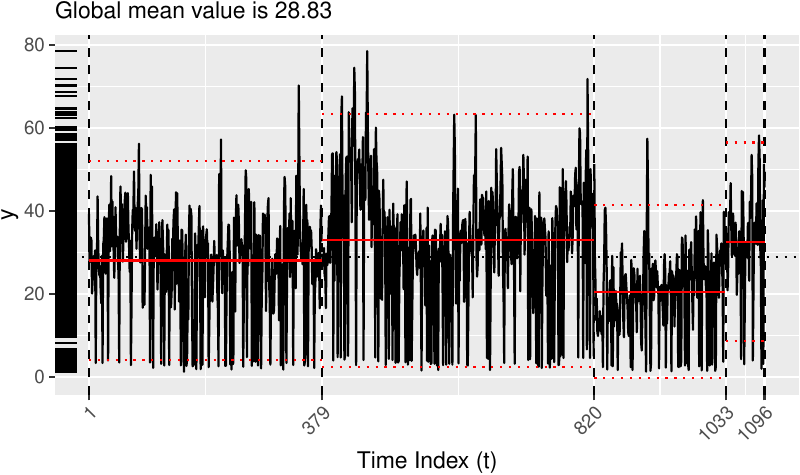}}

}

\subcaption{\label{fig-bog-compare-1}Coen's algorithm.}

\end{minipage}%
\begin{minipage}{0.50\linewidth}

\centering{

\pandocbounded{\includegraphics[keepaspectratio]{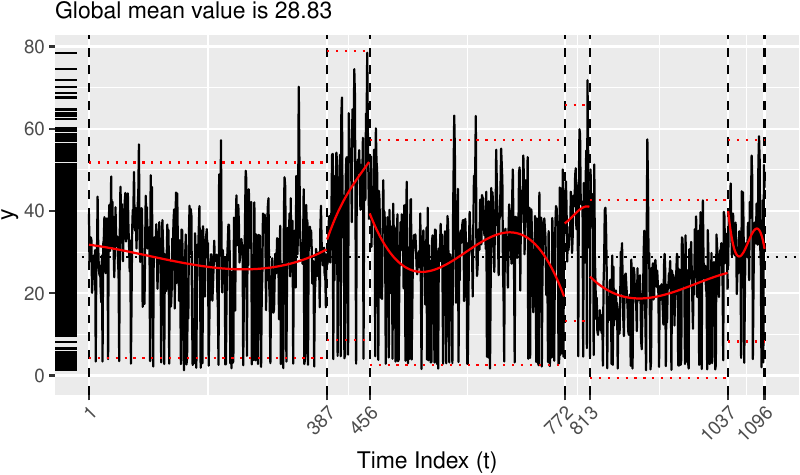}}

}

\subcaption{\label{fig-bog-compare-2}A genetic algorithm with a
polynomial trend model.}

\end{minipage}%

\caption{\label{fig-bog-compare}Changepoint sets for two genetic
algorithms applied to the Bogotá particulate matter data. Both pick up
on the COVID-19-related changepoint in March of 2020. The latter
includes 3rd degree polynomial trends in each region.}

\end{figure}

A more thorough analysis might use the \texttt{decompose()} or
\texttt{stl()} functions from the {stats} package to unravel the
seasonality of the particulate matter data. Instead, for the purposes of
simplicity here we simply experiment with different models. Setting the
\texttt{method} argument of \texttt{segment()} to \texttt{ga} allows us
to employ a custom genetic algorithm for which we can specify a model
via the \texttt{model\_fn} argument. In this case, we use the
\texttt{fit\_lmshift()} function (see Section~\ref{sec-model-fits}) to
fit a third-degree polynomial (note the \texttt{model\_fn\_args}
argument) to each region. Here, we use the \texttt{MDL()} penalty
function and the \texttt{log\_gabin\_population()} function. Both models
in Figure~\ref{fig-bog-compare} pick up on the COVID-19-related
changepoint. The piecewise constant model in
Figure~\ref{fig-bog-compare-1} ignores the seasonality present in the
data, while the Figure~\ref{fig-bog-compare-2} is undoubtedly overfit.

\begin{Shaded}
\begin{Highlighting}[]
\NormalTok{bog\_ga }\OtherTok{\textless{}{-}}\NormalTok{ bogota\_pm }\SpecialCharTok{|\textgreater{}}
  \FunctionTok{segment}\NormalTok{(}
    \AttributeTok{method =} \StringTok{"ga"}\NormalTok{, }\AttributeTok{model\_fn =}\NormalTok{ fit\_lmshift, }
    \AttributeTok{model\_fn\_args =} \FunctionTok{list}\NormalTok{(}\AttributeTok{deg\_poly =} \DecValTok{3}\NormalTok{), }
    \AttributeTok{penalty\_fn =}\NormalTok{ MDL, }
    \AttributeTok{population =} \FunctionTok{log\_gabin\_population}\NormalTok{(bogota\_pm), }
    \AttributeTok{maxiter =} \DecValTok{500}\NormalTok{, }\AttributeTok{run =} \DecValTok{100}
\NormalTok{  )}
\end{Highlighting}
\end{Shaded}

Note also that, without running the genetic algorithm again, we can
determine that adding AR(1) lagged errors to our model results in an
even lower MDL score.

\begin{Shaded}
\begin{Highlighting}[]
\FunctionTok{fitness}\NormalTok{(bog\_ga)}
\end{Highlighting}
\end{Shaded}

\begin{verbatim}
     MDL 
8828.881 
\end{verbatim}

\begin{Shaded}
\begin{Highlighting}[]
\NormalTok{bogota\_pm }\SpecialCharTok{|\textgreater{}}
  \FunctionTok{fit\_trendshift\_ar1}\NormalTok{(}\AttributeTok{tau =} \FunctionTok{changepoints}\NormalTok{(bog\_ga)) }\SpecialCharTok{|\textgreater{}}
  \FunctionTok{MDL}\NormalTok{()}
\end{Highlighting}
\end{Shaded}

\begin{verbatim}
[1] 8804.22
\end{verbatim}

\begin{figure}

\begin{minipage}{0.50\linewidth}

\centering{

\includegraphics[width=8in,height=\textheight,keepaspectratio]{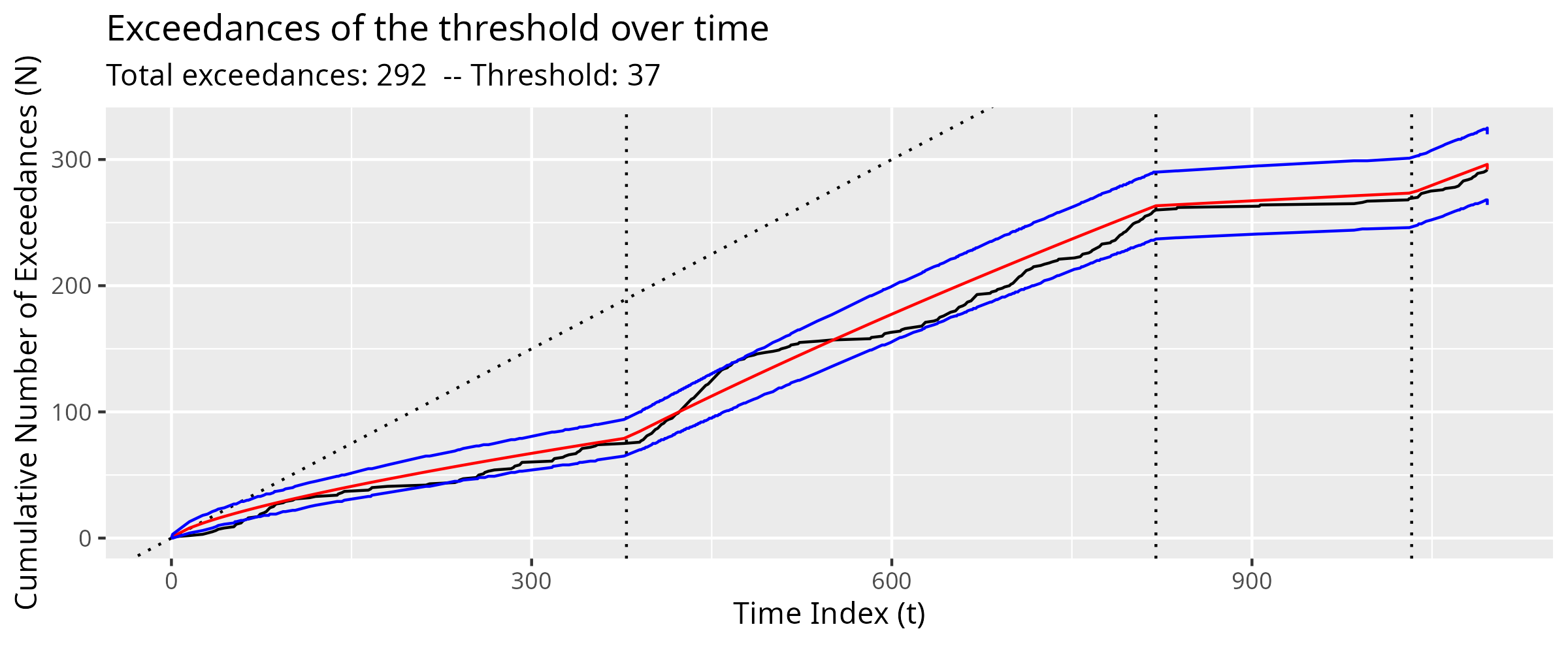}

}

\subcaption{\label{fig-diagnose-1}The \texttt{diagnose()} function
applied to a model object of class \texttt{nhpp} (Coen's algorithm).}

\end{minipage}%
\begin{minipage}{0.50\linewidth}

\centering{

\includegraphics[width=8in,height=\textheight,keepaspectratio]{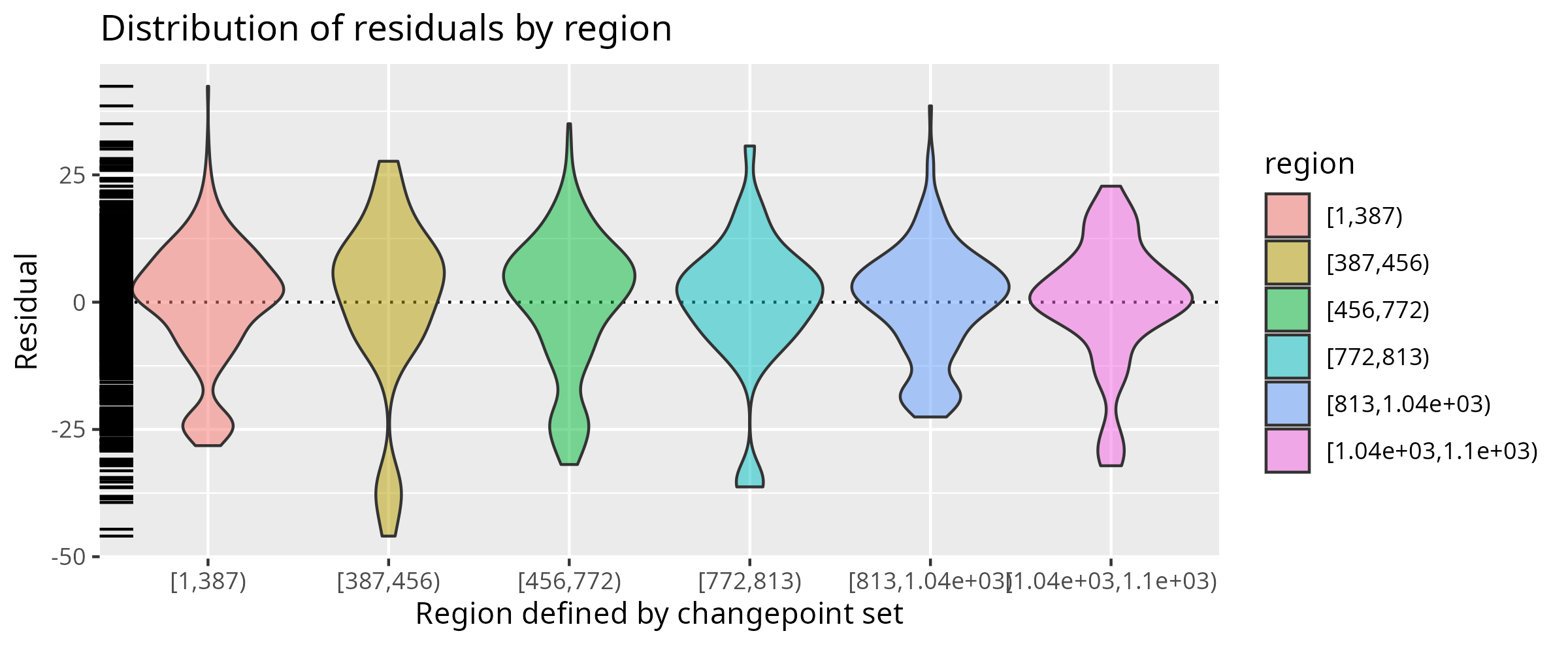}

}

\subcaption{\label{fig-diagnose-2}The \texttt{diagnose()} function
applied to a model object of class \texttt{mod\_cpt} (a genetic
algorithm with a polynomial trend model).}

\end{minipage}%

\caption{\label{fig-diagnose}Diagnostic plot for two algorithms applied
to the Bogotá particulate matter data. Note how the diagnostic plots
differ based on the model type.}

\end{figure}

The \texttt{diagnose()} function provides an informative visual
understanding of the quality of the fit of the model, which may vary
depending on the type of model. In Figure~\ref{fig-diagnose}, note how
Figure~\ref{fig-diagnose-2} summarizes the distribution of the
residuals, whereas Figure~\ref{fig-diagnose-1} shows the growth of the
cumulative number of exceedances (see Section~\ref{sec-nhpp}) relative
to the expected growth based on the NHPP model, with the blue lines
showing a 95\% confidence interval.

\section{Architecture of tidychangepoint}\label{sec-tidychangepoint}

The {tidychangepoint} package implements a simple, flexible structure
for working with a myriad of changepoint detection algorithms.
{tidychangepoint} makes extensive use of the S3 object-oriented class
system in R. In addition to creating a few new generic functions, it
provides methods for a handful of generic functions from other packages
(most notably {stats} and {broom}) for three new classes of objects
(\texttt{tidycpt}, \texttt{seg\_cpt}, and \texttt{mod\_cpt}), as well as
analogous methods for existing objects created by other R packages for
changepoint detection (e.g., {changepoint}). We provide more details in
this section.

Every \texttt{tidycpt} object is created by a call to the
\texttt{segment()} function. As a verb, ``segment'' describes the action
of splitting a time series into regions, and its usage accords with the
{tidyverse} convention of using verbs as function names. The
\texttt{method} argument indicates the algorithm we want to use, and any
subsequent arguments (e.g., \texttt{penalty}) are passed to the
corresponding function. Here, we pass the \texttt{penalty} argument to
force \texttt{changepoint::cpt.meanvar()} to use the BIC penalty
function.

\begin{Shaded}
\begin{Highlighting}[]
\NormalTok{mlb\_pelt\_meanvar\_bic }\OtherTok{\textless{}{-}}\NormalTok{ mlb\_bavg }\SpecialCharTok{|\textgreater{}}
  \FunctionTok{segment}\NormalTok{(}\AttributeTok{method =} \StringTok{"pelt"}\NormalTok{, }\AttributeTok{model\_fn =}\NormalTok{ fit\_meanvar, }\AttributeTok{penalty =} \StringTok{"BIC"}\NormalTok{)}
\FunctionTok{class}\NormalTok{(mlb\_pelt\_meanvar\_bic)}
\end{Highlighting}
\end{Shaded}

\begin{verbatim}
[1] "tidycpt"
\end{verbatim}

\subsection{Object structure}\label{object-structure}

Every \texttt{tidycpt} object contains three sub-objects:

\begin{itemize}
\tightlist
\item
  \texttt{segmenter}: An object that stores information about an optimal
  set of changepoints found by an algorithm. A \texttt{segmenter} object
  could be a \texttt{cpt} object returned by the \texttt{cpt.meanvar()}
  function from {changepoint}, a \texttt{ga} object returned by the
  \texttt{ga()} function from {GA}, or in principle, any object that
  provides methods for a few simple generic functions outlined in
  Section~\ref{sec-extending}.
\item
  \texttt{model}: A model object inheriting from \texttt{mod\_cpt}, an
  internal class for representing model objects. Model objects are
  created by model-fitting functions, all of whose names start with
  \texttt{fit\_} (see Section~\ref{sec-models}). The \texttt{model} of a
  \texttt{tidycpt} object is the model object returned by the
  \texttt{fit\_*()} function that corresponds to the one used by the
  segmenter. Because the models described in Section~\ref{sec-models}
  can be fit to any time series based solely on a specified set of
  changepoints, the information in this object does \emph{not} depend on
  the algorithm used to segment the times series: it only depends on the
  set of changepoints returned by the algorithm.
\item
  \texttt{elapsed\_time}: the clock time that elapsed while the
  algorithm was running.
\end{itemize}

\subsection{Methods available}\label{sec-methods}

\texttt{tidycpt} objects implement methods for the generic functions
\texttt{as.model()}, \texttt{as.segmenter()}, \texttt{as.ts()},
\texttt{changepoints()}, \texttt{diagnose()}, \texttt{fitness()},
\texttt{model\_name()}, \texttt{regions()}, \texttt{plot()}, and
\texttt{summary()}, as well as three generic functions from the {broom}
package: \texttt{augment()}, \texttt{tidy()}, and \texttt{glance()}.

\begin{Shaded}
\begin{Highlighting}[]
\FunctionTok{methods}\NormalTok{(}\AttributeTok{class =} \StringTok{"tidycpt"}\NormalTok{)}
\end{Highlighting}
\end{Shaded}

\begin{verbatim}
 [1] as.model     as.segmenter as.ts        augment      changepoints
 [6] diagnose     fitness      glance       model_name   plot        
[11] print        regions      summary      tidy        
see '?methods' for accessing help and source code
\end{verbatim}

For the most part, methods for \texttt{tidycpt} objects typically defer
to the methods defined for either the \texttt{segmenter} or the
\texttt{model}, depending on the user's likely intention. To that end,
both \emph{segmenters} and \emph{models} implement methods for the
generic functions \texttt{as.ts()}, \texttt{changepoints()},
\texttt{model\_name()}, \texttt{nobs()}, \texttt{plot()}, and
\texttt{print()}.

\begin{verbatim}
[1] "as.ts"        "changepoints" "model_name"   "nobs"         "plot"        
[6] "print"       
\end{verbatim}

The \texttt{changepoints()} function returns the \emph{time indices} of
the changepoint set, unless the \texttt{use\_labels} argument is set to
\texttt{TRUE}.

\begin{Shaded}
\begin{Highlighting}[]
\FunctionTok{changepoints}\NormalTok{(mlb\_pelt\_meanvar\_bic)}
\end{Highlighting}
\end{Shaded}

\begin{verbatim}
[1] 15 18 52 87
\end{verbatim}

\begin{Shaded}
\begin{Highlighting}[]
\FunctionTok{changepoints}\NormalTok{(mlb\_pelt\_meanvar\_bic, }\AttributeTok{use\_labels =} \ConstantTok{TRUE}\NormalTok{) }\SpecialCharTok{|\textgreater{}}
  \FunctionTok{as\_year}\NormalTok{()}
\end{Highlighting}
\end{Shaded}

\begin{verbatim}
[1] "1939" "1942" "1976" "2011"
\end{verbatim}

\begin{figure}[H]

\centering{

\pandocbounded{\includegraphics[keepaspectratio]{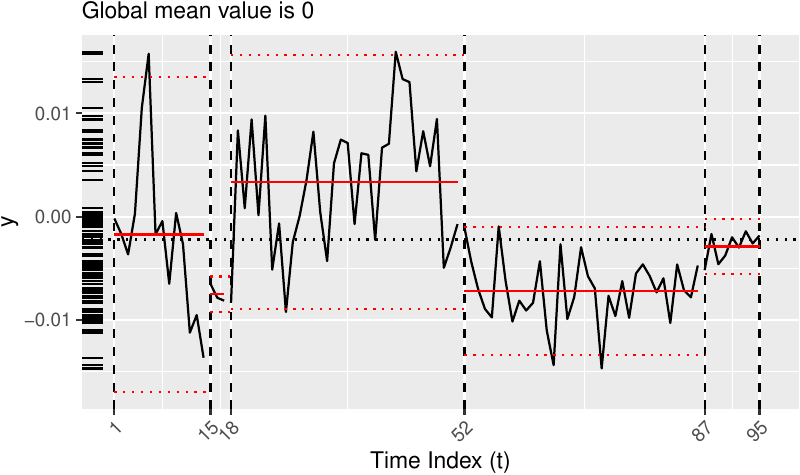}}

}

\caption{\label{fig-mlb-segmented}The MLB time series, as segmented by
the PELT algorithm using the normal meanvar model and the BIC penalty.}

\end{figure}

For example, the \texttt{plot()} method for \texttt{tidycpt} simply
calls the \texttt{plot()} method for the \texttt{model} of the
\texttt{tidycpt} object. The plot shown in
Figure~\ref{fig-mlb-segmented} uses {ggplot2} to illustrate how the
proposed changepoint set segments the time series.

\phantomsection\label{cell-fig-mlb-segmented}
\begin{Shaded}
\begin{Highlighting}[]
\FunctionTok{plot}\NormalTok{(mlb\_pelt\_meanvar\_bic)}
\end{Highlighting}
\end{Shaded}

Changepoint detection models in {tidychangepoint} follow the design
interface of the {broom} package. Therefore, \texttt{augment()},
\texttt{tidy()}, and \texttt{glance()} methods exist for
\texttt{mod\_cpt} objects.

\begin{itemize}
\tightlist
\item
  \texttt{augment()} returns a \texttt{tsibble} (E. Wang, Cook, and
  Hyndman 2020) that is grouped according to the regions defined by the
  changepoint set. This object class is appropriate for this use case,
  since the grouping by changepoint region is natural in this context,
  and a \texttt{tsibble} is a grouped \texttt{tibble} that is aware of
  its indexing variable. Users can of course undo the grouping with
  \texttt{dplyr::ungroup()} or coerce to a regular \texttt{tibble} with
  \texttt{tibble::as\_tibble()}.
\end{itemize}

\begin{Shaded}
\begin{Highlighting}[]
\FunctionTok{augment}\NormalTok{(mlb\_pelt\_meanvar\_bic)}
\end{Highlighting}
\end{Shaded}

\begin{verbatim}
# A tsibble: 95 x 5 [1]
# Groups:    region [5]
   index         y region  .fitted     .resid
   <int>     <dbl> <fct>     <dbl>      <dbl>
 1     1 -0.000173 [1,15) -0.00173  0.00156  
 2     2 -0.00166  [1,15) -0.00173  0.0000703
 3     3 -0.00366  [1,15) -0.00173 -0.00192  
 4     4  0.000285 [1,15) -0.00173  0.00202  
 5     5  0.0105   [1,15) -0.00173  0.0123   
 6     6  0.0158   [1,15) -0.00173  0.0175   
 7     7 -0.00171  [1,15) -0.00173  0.0000243
 8     8 -0.000439 [1,15) -0.00173  0.00130  
 9     9 -0.00650  [1,15) -0.00173 -0.00477  
10    10  0.000354 [1,15) -0.00173  0.00209  
# i 85 more rows
\end{verbatim}

\begin{itemize}
\tightlist
\item
  \texttt{tidy()} returns a \texttt{tibble} (Müller and Wickham 2025)
  that provides summary statistics for each region. These include any
  parameters that were fit, which are prefixed in the output by
  \texttt{param\_}.
\end{itemize}

\begin{Shaded}
\begin{Highlighting}[]
\FunctionTok{tidy}\NormalTok{(mlb\_pelt\_meanvar\_bic)}
\end{Highlighting}
\end{Shaded}

\begin{verbatim}
# A tibble: 5 x 10
  region  num_obs      min       max     mean       sd begin   end param_mu
  <chr>     <int>    <dbl>     <dbl>    <dbl>    <dbl> <dbl> <dbl>    <dbl>
1 [1,15)       14 -0.0137   0.0158   -0.00173 0.00779      1    15 -0.00173
2 [15,18)       3 -0.00813 -0.00650  -0.00750 0.000872    15    18 -0.00750
3 [18,52)      34 -0.00923  0.0160    0.00335 0.00628     18    52  0.00335
4 [52,87)      35 -0.0147  -0.000943 -0.00719 0.00315     52    87 -0.00719
5 [87,96)       9 -0.00518 -0.00142  -0.00290 0.00135     87    96 -0.00290
# i 1 more variable: param_sigma_hatsq <dbl>
\end{verbatim}

\begin{itemize}
\tightlist
\item
  \texttt{glance()} returns a one-row \texttt{tibble} that provides
  summary statistics for the algorithm. This includes the fitness, which
  is the value of the penalized objective function (see
  Section~\ref{sec-penalty}) that was used.
\end{itemize}

\begin{Shaded}
\begin{Highlighting}[]
\FunctionTok{glance}\NormalTok{(mlb\_pelt\_meanvar\_bic)}
\end{Highlighting}
\end{Shaded}

\begin{verbatim}
# A tibble: 1 x 8
  pkg      version algorithm seg_params model_name criteria fitness elapsed_time
  <chr>    <pckg_> <chr>     <list>     <chr>      <chr>      <dbl> <drtn>      
1 changep~ 2.3     PELT      <list [1]> meanvar    BIC        -782. 0.01 secs   
\end{verbatim}

\section{Segmenters}\label{sec-alg}

In the example above, the \texttt{segmenter} is of class \texttt{cpt},
because \texttt{segment()} simply wraps the \texttt{cpt.meanvar()}
function from the {changepoint} package.

\begin{Shaded}
\begin{Highlighting}[]
\NormalTok{mlb\_pelt\_meanvar\_bic }\SpecialCharTok{|\textgreater{}}
  \FunctionTok{as.segmenter}\NormalTok{() }\SpecialCharTok{|\textgreater{}}
  \FunctionTok{class}\NormalTok{()}
\end{Highlighting}
\end{Shaded}

\begin{verbatim}
[1] "cpt"
attr(,"package")
[1] "changepoint"
\end{verbatim}

In addition to the generic functions listed above, \emph{segmenters}
implement methods for the generic functions \texttt{as.seg\_cpt()},
\texttt{fitness()}, \texttt{model\_args()}, and \texttt{seg\_params()}.

\begin{verbatim}
[1] "as.seg_cpt" "fitness"    "model_args" "seg_params"
\end{verbatim}

Note that while {tidychangepoint} uses only S3 classes, segmenters (such
as those of class \texttt{cpt}) may belong to S4 classes. This is one
reason why \texttt{as.seg\_cpt()}, which coverts a segmenter of any
class to a \texttt{mod\_cpt} object, is necessary. \texttt{fitness()}
returns a named vector with the type and value of the penalized
objective function used by the segmenting algorithm.

\begin{Shaded}
\begin{Highlighting}[]
\FunctionTok{fitness}\NormalTok{(mlb\_pelt\_meanvar\_bic)}
\end{Highlighting}
\end{Shaded}

\begin{verbatim}
      BIC 
-782.1165 
\end{verbatim}

\subsection{Segmenting algorithms provided}\label{sec-seg-algs}

Table~\ref{tbl-coverage} shows the values passable to the
\texttt{method} argument of the \texttt{segment()} function and what
they do. Currently, \texttt{segment()} wraps the following algorithms:

\begin{itemize}
\tightlist
\item
  PELT and Binary Segmentation as implemented by {changepoint} (Killick
  2024)
\item
  Wild Binary Segmentation as implemented by {wbs} (Baranowski and
  Fryzlewicz 2024)
\item
  Segmented regression models as implemented by {segmented} (V. M. R.
  Muggeo 2025)
\item
  A variety of genetic algorithms as implemented by {GA} (Scrucca 2024)
  and {changepointGA} (M. Li and Lu 2025). Specific variants include
  Shi's algorithm (Shi et al. 2022) and Coen's algorithm (Suárez-Sierra,
  Coen, and Taimal 2023).
\item
  Trivial algorithms for no changepoints, manually input changepoints,
  and randomly selected changepoints
\end{itemize}

\begin{longtable}[]{@{}
  >{\raggedright\arraybackslash}p{(\linewidth - 8\tabcolsep) * \real{0.1100}}
  >{\raggedright\arraybackslash}p{(\linewidth - 8\tabcolsep) * \real{0.1900}}
  >{\raggedright\arraybackslash}p{(\linewidth - 8\tabcolsep) * \real{0.1400}}
  >{\raggedright\arraybackslash}p{(\linewidth - 8\tabcolsep) * \real{0.2600}}
  >{\raggedright\arraybackslash}p{(\linewidth - 8\tabcolsep) * \real{0.3000}}@{}}

\caption{\label{tbl-coverage}Segmenting methods provided by
tidychangepoint.}

\tabularnewline

\toprule\noalign{}
\begin{minipage}[b]{\linewidth}\raggedright
Method
\end{minipage} & \begin{minipage}[b]{\linewidth}\raggedright
Pkg
\end{minipage} & \begin{minipage}[b]{\linewidth}\raggedright
Segmenter Class
\end{minipage} & \begin{minipage}[b]{\linewidth}\raggedright
Helper
\end{minipage} & \begin{minipage}[b]{\linewidth}\raggedright
Wraps
\end{minipage} \\
\midrule\noalign{}
\endhead
\bottomrule\noalign{}
\endlastfoot
binseg & changepoint & cpt & NA & changepoint::cpt.meanvar() \\
coen & tidychangepoint & seg\_basket & segment\_coen() & NA \\
cptga & changepointGA & tidycptga & segment\_cptga() &
changepointGA::cptga() \\
ga & GA & tidyga & segment\_ga() & GA::ga() \\
ga-coen & GA & tidyga & segment\_ga\_coen() & segment\_ga() \\
ga-shi & GA & tidyga & segment\_ga\_shi() & segment\_ga() \\
manual & tidychangepoint & seg\_cpt & segment\_manual() & NA \\
null & tidychangepoint & seg\_cpt & segment\_manual() & NA \\
pelt & changepoint & cpt & segment\_pelt() & changepoint::cpt.mean() or
changepoint::cpt.meanvar() \\
random & GA & tidyga & segment\_ga\_random() & segment\_ga() \\
segmented & segmented & segmented & NA & segmented::segmented() \\
segneigh & changepoint & cpt & NA & changepoint::cpt.meanvar() \\
single-best & changepoint & cpt & NA & changepoint::cpt.meanvar() \\
strucchange & strucchange & breakpointsfull & NA &
strucchange::breakpoints() \\
wbs & wbs & wbs & NA & wbs::wbs() \\

\end{longtable}

In particular, setting the \texttt{method} argument to the
\texttt{segment()} function to \texttt{ga} allows the user to specify
any of the model-fitting functions described in
Section~\ref{sec-model-fits} (or a user-defined model-fitting function
as in Section~\ref{sec-extending-models}) as well as any of the
penalized objective functions in Section~\ref{sec-penalty}. While these
algorithms can be slow, they are quite flexible.

\subsection{Extending tidychangepoint to include new
algorithms}\label{sec-extending}

In order for a segmenting algorithm to work with {tidychangepoint} the
class of the resulting object must implement, at a minimum, methods for
the following generic functions:

\begin{itemize}
\tightlist
\item
  From {stats}: \texttt{as.ts()}, and \texttt{nobs()}. These are needed
  perform basic operations on the time series.
\item
  From {tidychangepoint}: \texttt{changepoints()} must return a single
  integer vector with the best changepoint set as defined by the
  algorithm. \texttt{fitness()} must return a named double vector with
  the appropriately named value of the penalized objective function.
  \texttt{as.segmenter()} dumps the contents of the object into a
  \texttt{mod\_cpt} object. \texttt{model\_name()},
  \texttt{model\_args()}, and \texttt{seg\_params()} provide information
  about the kind of model that was fit, and any arguments passed to the
  segmenting function.
\end{itemize}

These compatibility layers are generally not difficult or time-consuming
to write.

\subsection{Spotlight: Coen's algorithm}\label{sec-coen}

The goal of Coen's algorithm (Suárez-Sierra, Coen, and Taimal 2023) is
to find the optimal set of changepoints for the NHPP model described in
Section~\ref{sec-nhpp}. That is, given a time series \(y\), find the
changepoint set \(\tau\) that minimizes the value of
\(BMDL(\tau, NHPP(y | \hat{\theta}_\tau))\). To find good candidates,
this genetic algorithm starts with a randomly selected set of
\texttt{popSize} changepoint sets, and then iteratively ``mates'' the
best (according to the BMDL penalty function, see
Section~\ref{sec-bmdl}) pairs of these changepoint sets to recover a new
``generation'' of \texttt{popSize} changepoint sets. This process is
repeated until a stopping criteria is met (most simply, \texttt{maxiter}
times). Thus, a constant number (\texttt{popSize} \(\times\)
\texttt{maxiter}) possible changepoint sets are considered, and the one
with the lowest BMDL score is selected.

Coen's algorithm is implemented in {tidychangepoint} via the
\texttt{segment()} function with the \texttt{method} argument set to
\texttt{ga-coen}. As the code below reveals, this is just a special case
of the \texttt{segment\_ga()} function that wraps \texttt{GA::ga()}.
Note that the \texttt{model\_fn} argument is set to \texttt{fit\_nhpp}
and the \texttt{penalty\_fn} argument is set to \texttt{BMDL}. The
running time of the back-end function \texttt{GA::ga()} is sensitive to
the size of the changepoint sets considered, especially in the first
generation, and thus we use the \texttt{population} argument to inform
the selection of the first generation (see below). Coen's algorithm runs
in about the same time as a naïve algorithm that randomly selects the
same number of changepoints, but produces changepoint sets with
significantly better BMDL scores.

\begin{Shaded}
\begin{Highlighting}[]
\NormalTok{segment\_ga\_coen}
\end{Highlighting}
\end{Shaded}

\begin{verbatim}
function(x, ...) {
  segment_ga(
    x, model_fn = fit_nhpp, penalty_fn = BMDL, 
    population = build_gabin_population(x), popSize = 50, ...
  )
}
<bytecode: 0x6152682dfe80>
<environment: namespace:tidychangepoint>
\end{verbatim}

By default, the function \texttt{GA::gabin\_Population()} selects
candidate changepoints uniformly at random with probability 0.5, leading
to candidate changepoint sets of size \(n/2\), on average. These
candidate changepoint sets are usually poor (since \(n/2\) changepoints
is ludicrous), and we observe much better and faster performance by
seeding the first generation with smaller candidate changepoint sets. To
this end, the \texttt{build\_gabin\_population()} function runs several
fast algorithms (i.e., PELT, Wild Binary Segmentation, etc.) and sets
the initial probability of being selected to three times the average
size of the changepoint set found by these algorithms. This results in
much more rapid covergence around the optimal changepoint set.
Alternatively, {tidychangepoint} also provides the
\texttt{log\_gabin\_population()} function, which sets the initial
probability to \(\ln{n} / n\) (see Section~\ref{sec-benchmarking}).

\section{Models}\label{sec-models}

All \texttt{model} objects are created by calls to one of the
model-fitting functions listed in Section~\ref{sec-model-fits}, whose
name begins with \texttt{fit\_}. These functions all inherit from the
class \texttt{fun\_cpt} and can be listed using \texttt{ls\_models()}.
All \texttt{model} objects inherit from the \texttt{mod\_cpt} base
class.

The \texttt{model} object in our example is created by
\texttt{fit\_meanvar()}, and is of class \texttt{mod\_cpt}.

\begin{Shaded}
\begin{Highlighting}[]
\NormalTok{mlb\_pelt\_meanvar\_bic }\SpecialCharTok{|\textgreater{}}
  \FunctionTok{as.model}\NormalTok{() }\SpecialCharTok{|\textgreater{}}
  \FunctionTok{str}\NormalTok{()}
\end{Highlighting}
\end{Shaded}

\begin{verbatim}
List of 6
 $ data         : Time-Series [1:95] from 1 to 95: -0.000173 -0.001663 -0.003656 0.000285 0.010521 ...
 $ tau          : int [1:4] 15 18 52 87
 $ region_params: tibble [5 x 3] (S3: tbl_df/tbl/data.frame)
  ..$ region           : chr [1:5] "[1,15)" "[15,18)" "[18,52)" "[52,87)" ...
  ..$ param_mu         : num [1:5] -0.00173 -0.0075 0.00335 -0.00719 -0.0029
  ..$ param_sigma_hatsq: Named num [1:5] 5.63e-05 5.07e-07 3.83e-05 9.67e-06 1.62e-06
  .. ..- attr(*, "names")= chr [1:5] "[1,15)" "[15,18)" "[18,52)" "[52,87)" ...
 $ model_params : NULL
 $ fitted_values: num [1:95] -0.00173 -0.00173 -0.00173 -0.00173 -0.00173 ...
 $ model_name   : chr "meanvar"
 - attr(*, "class")= chr "mod_cpt"
\end{verbatim}

In addition to the generic functions listed above, models implement
methods for the generic functions \texttt{coef()}, \texttt{fitted()},
\texttt{logLik()}, \texttt{plot()}, and \texttt{residuals()}, as well as
\texttt{augment()}, \texttt{tidy()}, and \texttt{glance()}.

\begin{verbatim}
 [1] "augment"   "coef"      "diagnose"  "fitted"    "glance"    "logLik"   
 [7] "regions"   "residuals" "summary"   "tidy"     
\end{verbatim}

Like other model objects that implement \texttt{glance()} methods, the
output from the \texttt{glance()} function summarizes the fit of the
model to the data.

\begin{Shaded}
\begin{Highlighting}[]
\NormalTok{mlb\_pelt\_meanvar\_bic }\SpecialCharTok{|\textgreater{}}
  \FunctionTok{as.model}\NormalTok{() }\SpecialCharTok{|\textgreater{}}
  \FunctionTok{glance}\NormalTok{()}
\end{Highlighting}
\end{Shaded}

\begin{verbatim}
# A tibble: 1 x 11
  pkg   version algorithm params num_cpts    rmse logLik   AIC   BIC  MBIC   MDL
  <chr> <pckg_> <chr>     <list>    <int>   <dbl>  <dbl> <dbl> <dbl> <dbl> <dbl>
1 tidy~ 1.0.3   meanvar   <NULL>        4 0.00507   367. -706. -671. -689. -674.
\end{verbatim}

\subsection{Model-fitting functions provided}\label{sec-model-fits}

Table~\ref{tbl-models} lists the model-fitting functions provided by
{tidychangepoint}, along with their parameters. Any of these functions
can be passed to the \texttt{model\_fn} argument of
\texttt{segment\_ga()} (see Section~\ref{sec-seg-algs}).

\begin{longtable}[]{@{}
  >{\raggedright\arraybackslash}p{(\linewidth - 6\tabcolsep) * \real{0.3200}}
  >{\raggedright\arraybackslash}p{(\linewidth - 6\tabcolsep) * \real{0.1400}}
  >{\raggedright\arraybackslash}p{(\linewidth - 6\tabcolsep) * \real{0.3000}}
  >{\raggedright\arraybackslash}p{(\linewidth - 6\tabcolsep) * \real{0.2400}}@{}}

\caption{\label{tbl-models}Model-fitting functions provided by
tidychangepoint. Unless otherwise noted, all models assume a normal
distribution and white noise errors. The suffix \texttt{\_ar1} indicates
autocorrelated AR(1) lagged errors (as decribed in Shi et al. (2022)),
where \(\phi\) is the measure of autocorrelation. See
Section~\ref{sec-nhpp} for a description of the non-homogeneous Poisson
process model used by \texttt{fit\_nhpp()}.}

\tabularnewline

\toprule\noalign{}
\begin{minipage}[b]{\linewidth}\raggedright
Function name
\end{minipage} & \begin{minipage}[b]{\linewidth}\raggedright
\(k\)
\end{minipage} & \begin{minipage}[b]{\linewidth}\raggedright
Parameters (\(\theta\))
\end{minipage} & \begin{minipage}[b]{\linewidth}\raggedright
Note
\end{minipage} \\
\midrule\noalign{}
\endhead
\bottomrule\noalign{}
\endlastfoot
\texttt{fit\_meanshift\_norm()} & \(m+2\) &
\(\sigma^2, \mu_0, \ldots, \mu_m\) & Normal \\
\texttt{fit\_meanshift\_lnorm()} & \(m+2\) &
\(\sigma^2, \mu_0, \ldots, \mu_m\) & log-normal \\
\texttt{fit\_meanshift\_pois()} & \(m+2\) &
\(\sigma^2, \mu_0, \ldots, \mu_m\) & Poisson \\
\texttt{fit\_meanshift\_norm\_ar1()} & \(m+3\) &
\(\sigma^2, \phi, \mu_0, \ldots, \mu_m\) & \\
\texttt{fit\_trendshift()} & \(2m + 3\) &
\(\sigma^2, \mu_0, \ldots, \mu_m, \beta_0, \ldots, \beta_m\) & special
case of \texttt{fit\_lmshift()} with \(p = 1\) \\
\texttt{fit\_trendshift\_ar1()} & \(2m + 4\) &
\(\sigma^2, \phi, \mu_0, \ldots, \mu_m, \beta_0, \ldots, \beta_m\) & \\
\texttt{fit\_lmshift()} & \(p(m + 1) + 1\) &
\(\sigma^2, \beta_{00}, \ldots, \beta_{0m}, \beta_{10}, \ldots, \beta_{pm}\)
& polynomial\footnote{Unlike \texttt{fit\_meanshift\_norm()},
  \texttt{fit\_lmshift()} function uses the \texttt{lm()} function from
  the {stats}, which provides flexibility at the expense of speed
  (\texttt{fit\_meanshift\_norm()} is faster).} of degree \(p\) \\
\texttt{fit\_lmshift\_ar1()} & \(p(m + 1) + 1\) &
\(\sigma^2, \beta_{00}, \ldots, \beta_{0m}, \beta_{10}, \ldots, \beta_{pm}\)
& \\
\texttt{fit\_meanvar()} & \(2m + 2\) &
\(\sigma_0^2, \ldots, \sigma_m^2, \mu_0, \ldots, \mu_m\) & \\
\texttt{fit\_nhpp()} & \(2m+2\) &
\(\alpha_0, \ldots, \alpha_m, \beta_0, \ldots, \beta_m\) & \\

\end{longtable}

\subsection{Extending tidychangepoint to include new
models}\label{sec-extending-models}

Users can create their own model-fitting functions. The names of these
functions should start with \texttt{fit\_} for consistency, and they
must be registered with a call to \texttt{fun\_cpt()}. The first
argument \texttt{x} must be the time series and the second argument
\texttt{tau} must be a subset of the indices of \texttt{x}. Any
subsequent arguments can be passed through the dots. Every
\texttt{fit\_} function returns an object of class \texttt{mod\_cpt},
which can be created with the arguments shown below.

\begin{Shaded}
\begin{Highlighting}[]
\FunctionTok{args}\NormalTok{(new\_mod\_cpt)}
\end{Highlighting}
\end{Shaded}

\begin{verbatim}
function (x = numeric(), tau = integer(), region_params = tibble::tibble(), 
    model_params = double(), fitted_values = double(), model_name = character(), 
    ...) 
NULL
\end{verbatim}

\subsection{Spotlight: Non-homogenous Poisson process
model}\label{sec-nhpp}

Having already described the meanshift model in Section~\ref{sec-intro},
we now describe mathematically the NHPP model used in Coen's algorithm
(see Section~\ref{sec-coen}) and created by the \texttt{fit\_nhpp()}
function listed in Section~\ref{sec-model-fits}.

Let \(z_w(y) = \{t : y_t > w\}\) be a subset of the \emph{indices} of
the original times series \(y\), for which the observation \(y_t\)
exceeds some threshold \(w\), which by default is the empirical mean
\(\bar{y}\). This new time series \(z_w(y)\) is called the
\emph{exceedances} of the original time series \(y\). Following
Suárez-Sierra, Coen, and Taimal (2023), we model the time series
\(z_w(y)\) as a non-homogenous Poisson process (NHPP), using a Weibull
distribution parameterized by shape (\(\alpha\)) and scale (\(\beta\)).
In a Bayesian setting, those parameters follow a Gamma distribution with
their appropriate hyperparameters. Thus, we assume that there exists a
set of parameters
\(\theta = (\alpha_0 \ldots \alpha_m, \beta_0, \ldots, \beta_m)\) for
which the exceedances \(z_w(y)\) of the original time series \(y\) are
modeled accurately as a non-homogenous Poisson process.

The best-fit parameters are determined by maximizing the log posterior
function described below: \[
  \ln{ g(\theta_\tau | y ) } \propto \ln{ L_{NHPP}(y | \theta_\tau) } + \ln{ g(\theta_\tau) } \,,
\] where \(L\) is the likelihood function and \(g\) is the prior
probability distribution function. Closed-form expressions for each of
these components for a Weibull distribution with two Gamma priors are
given in Suárez-Sierra, Coen, and Taimal (2023). The \texttt{tidy()}
function displays the values of the fitted parameter values, or one can
pull the \texttt{region\_params} object directly from the \texttt{model}
object.

\begin{Shaded}
\begin{Highlighting}[]
\NormalTok{bog\_coen }\SpecialCharTok{|\textgreater{}}
  \FunctionTok{as.model}\NormalTok{() }\SpecialCharTok{|\textgreater{}}
  \FunctionTok{pluck}\NormalTok{(}\StringTok{"region\_params"}\NormalTok{)}
\end{Highlighting}
\end{Shaded}

\begin{verbatim}
# A tibble: 4 x 5
  region             param_alpha param_beta logPost logLik
  <chr>                    <dbl>      <dbl>   <dbl>  <dbl>
1 [1,379)                  0.711     0.808   -202.  -198. 
2 [379,820)                0.661     0.0762  -346.  -344. 
3 [820,1.03e+03)           0.484     0.0831   -42.6  -40.6
4 [1.03e+03,1.1e+03)       0.668     0.0788   -48.4  -46.6
\end{verbatim}

\section{Penalized objective functions}\label{sec-penalty}

We call the function \(f\) a \emph{penalized objective
function}\footnote{Occasionally, we may abuse terminology by referring
  to \(f\) as a \emph{penalty function}, but technically, \(P_f\) is the
  penalty function.} because it comes in the form \[
  f(\tau, M(y|\theta_\tau)) = \underbrace{P_f(\tau, n)}_{penalty} - 2 \ln{ \underbrace{L_M(y| \theta_\tau)}_{likelihood}} \,,
\] where \(P_f\) is a function that guards against overfitting by
penalizing large changepoint sets---that is, it is a monotonically
increasing function of \(m\) (the number of changepoints). Recall that
\(\tau\) is the set of changepoints. The {stats} package (R Core Team
2025) provides generic methods for the well-known penalty functions
\texttt{AIC()} and \texttt{BIC()}, which work seamlessly on
{tidychangepoint} models because they all implement methods for
\texttt{logLik()}. In addition, {tidychangepoint} provides generic
functions and methods for four additional penalty functions:
\texttt{HQC()}, \texttt{MBIC()}, \texttt{MDL()}, and \texttt{BMDL()}.
For compatibility with {changepoint} we also provide the alias
\texttt{SIC()} for \texttt{BIC()}. {tidychangepoint} supports all of the
penalty functions supported by {changepoint}, with the exception of the
``Asymptotic'' and ``CROPS'' penalties, which are not fully
generalizable.

All penalty functions return 0 for the special case in which \(m = 0\).
For ease of explanation, let the vector \(\ell_j = \tau_{j+1} - \tau_j\)
for \(0 \leq j \leq m\) encode the lengths of the \(m+1\) regions
defined by the changepoint set \(\tau\).

\subsection{\texorpdfstring{Penalized objective functions supported in
{tidychangepoint}}{Penalized objective functions supported in tidychangepoint}}\label{penalized-objective-functions-supported-in-tidychangepoint}

\subsubsection{Hannan-Quinn Information
Criterion}\label{hannan-quinn-information-criterion}

The Hannan-Quinn information Criterion is a slight tweak on the BIC that
involves an iterated logarithm (Hannan and Quinn 1979):
\(P_{HQC}(\tau, n) = 2m \cdot \ln{\ln{n}}\).

\subsubsection{Modified Bayesian Information Criterion}\label{sec-mbic}

Following Zhang and Siegmund (2007) and S. Li and Lund (2012), we define
the MBIC as:

\[
  P_{MBIC}(\tau, n) = 3 m \ln{n} + \sum_{j=1}^{m+1} \ln{\frac{\ell_j}{n}} \,.
\]

\subsubsection{Minimum Descriptive Length}\label{sec-mdl}

As described in Shi et al. (2022) and S. Li and Lund (2012), we define
the MDL as:

\[
  P_{MDL}(\tau, n) = \frac{a(\theta_\tau)}{2} \cdot \sum_{j=0}^m \log{\ell_j} + 2 \ln{m} + \sum_{j=2}^m \ln{\tau_j} + \left( 2 + b(\theta_\tau) \right) \ln{n} \,,
\] where \(a(\theta)\) is the number of parameters in \(\theta\) that
are fit in each region, and \(b(\theta)\) is the number of parameters
fit to the model as a whole. For example, the in the meanshift model,
\(a(\theta) = 1\) to account for the \(\mu_i\)'s, and \(b(\theta) = 1\)
to account for \(\sigma^2\).

The MBIC and MDL differ from the penalties imposed by either the Akaike
Information Criterion (AIC) or the BIC in important ways. For example,
the penalty for the AIC depends just linearly on the number of
changepoints, while the BIC depends on the number of changepoints as
well as the length of the original times series. Conversely, the MDL
penalty depends not only on the number of changepoints, but the
\emph{values} of the changepoints. Because of this, changepoints far
from zero can have disproportionate impact.

\subsubsection{Bayesian Minimum Descriptive Length}\label{sec-bmdl}

The Bayesian Minimum Descriptive Length combines the MDL penalty
function with the log prior \(g\) for the best-fit parameters
\(\hat{\theta}\) in the NHPP model described in Section~\ref{sec-nhpp}.
Currently, the BMDL penalty can only be applied to the NHPP model.

The exact value of the BMDL is then: \[
  BMDL(\tau, NHPP(y | \hat{\theta}_\tau)) = P_{MDL}(\tau) - 2 \ln{ L_{NHPP}(y | \hat{\theta}_\tau) } - 2 \ln{ g(\hat{\theta}_\tau) } \,.
\]

\subsection{Extending tidychangepoint to include new penalty
functions}\label{sec-extending-penalties}

New penalty functions can be contributed to {tidychangepoint} by
defining a new generic function and implementing a method for the
\texttt{logLik} class. Penalty functions should return a \texttt{double}
vector of length one.

Note that similar to the \texttt{logLik()} method for \texttt{lm}, the
\texttt{logLik()} method for \texttt{mod\_cpt} embeds relevant
information into the object's attributes. These include the quantities
necessary to compute \texttt{AIC()}, \texttt{BIC()}, \texttt{MDL()},
etc.

\begin{Shaded}
\begin{Highlighting}[]
\NormalTok{mlb\_pelt\_meanvar\_bic }\SpecialCharTok{|\textgreater{}}
  \FunctionTok{as.model}\NormalTok{() }\SpecialCharTok{|\textgreater{}}
  \FunctionTok{logLik}\NormalTok{() }\SpecialCharTok{|\textgreater{}}
  \FunctionTok{unclass}\NormalTok{() }\SpecialCharTok{|\textgreater{}}
  \FunctionTok{str}\NormalTok{()}
\end{Highlighting}
\end{Shaded}

\begin{verbatim}
 num 367
 - attr(*, "num_params_per_region")= int 2
 - attr(*, "num_model_params")= int 0
 - attr(*, "df")= num 14
 - attr(*, "nobs")= int 95
 - attr(*, "tau")= int [1:4] 15 18 52 87
\end{verbatim}

\section{Results}\label{sec-results}

In this section we document {tidychangepoint}'s correctness and
computational costs.

\subsection{Correctness}\label{sec-correctness}

Since {tidychangepoint} relies on other packages to actually segment
time series via their respective changepoint detection algorithms, the
results returned by \texttt{segment()} are correct to the extent that
the dependent results are correct.

However, recall that the \texttt{model} component of a \texttt{tidycpt}
object is computed by our \texttt{fit\_*()} model-fitting functions,
\emph{after} the segmenting algorithm has returned its results. We show
below that these computations are accurate.

First, we match the model parameters returned by
\texttt{fit\_trendshift()} and \texttt{fit\_trendshift\_ar1()} with the
results for the Central England Temperature data (\texttt{CET}) reported
in rows 1--4 of Table 2 in Shi et al. (2022). Unfortunately, the CET
data been revised since the publication of Shi et al. (2022), so the
temperatures present in \texttt{CET} do not exactly match the
temperatures used in Shi et al. (2022) for all years. Accordingly, the
numbers in Table~\ref{tbl-shi-compare} do not match exactly, but they
are very close. However, when we perform this same analysis on the old
data (not shown here), the numbers match exactly.\footnote{Other results
  from Table 3 also match.}

\begin{Shaded}
\begin{Highlighting}[]
\NormalTok{cpts }\OtherTok{\textless{}{-}} \FunctionTok{c}\NormalTok{(}\DecValTok{1700}\NormalTok{, }\DecValTok{1739}\NormalTok{, }\DecValTok{1988}\NormalTok{)}
\NormalTok{ids }\OtherTok{\textless{}{-}} \FunctionTok{time2tau}\NormalTok{(cpts, }\FunctionTok{as\_year}\NormalTok{(}\FunctionTok{time}\NormalTok{(CET)))}
\NormalTok{trend\_wn }\OtherTok{\textless{}{-}}\NormalTok{ CET[}\StringTok{"/2020{-}01{-}01"}\NormalTok{] }\SpecialCharTok{|\textgreater{}} 
  \FunctionTok{fit\_trendshift}\NormalTok{(}\AttributeTok{tau =}\NormalTok{ ids)}
\NormalTok{trend\_ar1 }\OtherTok{\textless{}{-}}\NormalTok{ CET[}\StringTok{"/2020{-}01{-}01"}\NormalTok{] }\SpecialCharTok{|\textgreater{}} 
  \FunctionTok{fit\_trendshift\_ar1}\NormalTok{(}\AttributeTok{tau =}\NormalTok{ ids)}
\end{Highlighting}
\end{Shaded}

\begin{longtable}[]{@{}
  >{\raggedright\arraybackslash}p{(\linewidth - 10\tabcolsep) * \real{0.3333}}
  >{\raggedleft\arraybackslash}p{(\linewidth - 10\tabcolsep) * \real{0.2179}}
  >{\raggedleft\arraybackslash}p{(\linewidth - 10\tabcolsep) * \real{0.1026}}
  >{\raggedleft\arraybackslash}p{(\linewidth - 10\tabcolsep) * \real{0.0897}}
  >{\raggedleft\arraybackslash}p{(\linewidth - 10\tabcolsep) * \real{0.0897}}
  >{\raggedleft\arraybackslash}p{(\linewidth - 10\tabcolsep) * \real{0.1667}}@{}}

\caption{\label{tbl-shi-compare}Comparison of results reported by Shi et
al. (2022) to those reported by {tidychangepoint}. All discrepancies are
caused by revisions to the underlying CET data used.}

\tabularnewline

\toprule\noalign{}
\begin{minipage}[b]{\linewidth}\raggedright
Model
\end{minipage} & \begin{minipage}[b]{\linewidth}\raggedleft
\(\hat{\sigma}^2\)
\end{minipage} & \begin{minipage}[b]{\linewidth}\raggedleft
logLik
\end{minipage} & \begin{minipage}[b]{\linewidth}\raggedleft
BIC
\end{minipage} & \begin{minipage}[b]{\linewidth}\raggedleft
MDL
\end{minipage} & \begin{minipage}[b]{\linewidth}\raggedleft
\(\hat{\phi}\)
\end{minipage} \\
\midrule\noalign{}
\endhead
\bottomrule\noalign{}
\endlastfoot
Shi et al. (2022)
WN & 0.291 & -290.02 & 650.74 & 653.07 & NA \\
trend\_wn & 0.290 & -289.86 & 650.41 & 652.75 & NA \\
Shi et al. (2022)
AR(1) & 0.290 & -288.80 & 654.19 & 656.52 & 0.058 \\
trend\_ar1 & 0.290 & -288.50 & 653.59 & 655.94 & 0.055 \\

\end{longtable}

Second, in Table~\ref{tbl-param-compare} we compare the estimated
parameter values from the \texttt{model} component of a \texttt{tidycpt}
object (which are fit by \texttt{fit\_meanvar()}) to those from the
\texttt{segmenter} component (which are fit by \texttt{cpt.meanvar()})
using a simulated data set.

\begin{Shaded}
\begin{Highlighting}[]
\NormalTok{tidycpt }\OtherTok{\textless{}{-}}\NormalTok{ DataCPSim }\SpecialCharTok{|\textgreater{}}
  \FunctionTok{segment}\NormalTok{(}\AttributeTok{method =} \StringTok{"pelt"}\NormalTok{, }\AttributeTok{penalty =} \StringTok{"BIC"}\NormalTok{)}
\CommentTok{\# Results shown in Table...}
\CommentTok{\# tidycpt$segmenter@param.est}
\CommentTok{\# y$model$region\_params}
\end{Highlighting}
\end{Shaded}

\begin{longtable}[]{@{}
  >{\raggedright\arraybackslash}p{(\linewidth - 8\tabcolsep) * \real{0.1346}}
  >{\raggedleft\arraybackslash}p{(\linewidth - 8\tabcolsep) * \real{0.2115}}
  >{\raggedleft\arraybackslash}p{(\linewidth - 8\tabcolsep) * \real{0.2596}}
  >{\raggedleft\arraybackslash}p{(\linewidth - 8\tabcolsep) * \real{0.1731}}
  >{\raggedleft\arraybackslash}p{(\linewidth - 8\tabcolsep) * \real{0.2212}}@{}}

\caption{\label{tbl-param-compare}Comparison of parameter values
returned by \texttt{fit\_meanvar()} (\(\hat{\mu}_{tidycpt}\) and
\(\hat{\sigma}^2_{tidycpt}\)) to those of
\texttt{changepoint::cpt.meanvar()} (\(\hat{\mu}_{cpt}\) and
\(\hat{\sigma}^2_{cpt}\)). The discrepancies in the means are caused by
{tidychangepoint} using intervals open on the right, as opposed to the
left. Additional discrepancies in the variances are caused by
{changepoint}'s use of the population variance.}

\tabularnewline

\toprule\noalign{}
\begin{minipage}[b]{\linewidth}\raggedright
region
\end{minipage} & \begin{minipage}[b]{\linewidth}\raggedleft
\(\hat{\mu}_{tidycpt}\)
\end{minipage} & \begin{minipage}[b]{\linewidth}\raggedleft
\(\hat{\sigma}^2_{tidycpt}\)
\end{minipage} & \begin{minipage}[b]{\linewidth}\raggedleft
\(\hat{\mu}_{cpt}\)
\end{minipage} & \begin{minipage}[b]{\linewidth}\raggedleft
\(\hat{\sigma}^2_{cpt}\)
\end{minipage} \\
\midrule\noalign{}
\endhead
\bottomrule\noalign{}
\endlastfoot
{[}1,547) & 35.28 & 127.09 & 35.28 & 126.88 \\
{[}547,822) & 58.10 & 371.75 & 58.20 & 370.52 \\
{[}822,972) & 96.71 & 923.95 & 96.77 & 920.98 \\
{[}972,1.1e+03) & 155.85 & 2441.82 & 156.52 & 2405.97 \\

\end{longtable}

The tiny differences in the means are because {tidychangepoint} uses
intervals that are closed on the left and open on the right, whereas
{changepoint} does the opposite. The differences in the variances are
caused by the same issue and {changepoint}'s reporting of the population
variance instead of the sample variance.

\subsection{Benchmarking}\label{sec-benchmarking}

Since \texttt{segment()} wraps underlying functionality from other
packages, there is overhead created, both in terms of memory and speed.
Table~\ref{tbl-bench} shows the results of benchmarking
\texttt{segment()} against \texttt{changepoint::cpt.meanvar()}, which is
doing the underlying work in both cases. In this case,
\texttt{segment()} uses about twice as much memory and is about 8 times
slower than the underlying function. While that may seem like a lot,
we're talking about microseconds.

\begin{longtable}[]{@{}
  >{\raggedright\arraybackslash}p{(\linewidth - 16\tabcolsep) * \real{0.1928}}
  >{\raggedleft\arraybackslash}p{(\linewidth - 16\tabcolsep) * \real{0.1084}}
  >{\raggedleft\arraybackslash}p{(\linewidth - 16\tabcolsep) * \real{0.1084}}
  >{\raggedleft\arraybackslash}p{(\linewidth - 16\tabcolsep) * \real{0.1205}}
  >{\raggedleft\arraybackslash}p{(\linewidth - 16\tabcolsep) * \real{0.1205}}
  >{\raggedleft\arraybackslash}p{(\linewidth - 16\tabcolsep) * \real{0.0843}}
  >{\raggedleft\arraybackslash}p{(\linewidth - 16\tabcolsep) * \real{0.0723}}
  >{\raggedleft\arraybackslash}p{(\linewidth - 16\tabcolsep) * \real{0.0602}}
  >{\raggedleft\arraybackslash}p{(\linewidth - 16\tabcolsep) * \real{0.1325}}@{}}

\caption{\label{tbl-bench}Benchmarking}

\tabularnewline

\toprule\noalign{}
\begin{minipage}[b]{\linewidth}\raggedright
expression
\end{minipage} & \begin{minipage}[b]{\linewidth}\raggedleft
min
\end{minipage} & \begin{minipage}[b]{\linewidth}\raggedleft
median
\end{minipage} & \begin{minipage}[b]{\linewidth}\raggedleft
itr/sec
\end{minipage} & \begin{minipage}[b]{\linewidth}\raggedleft
mem\_alloc
\end{minipage} & \begin{minipage}[b]{\linewidth}\raggedleft
gc/sec
\end{minipage} & \begin{minipage}[b]{\linewidth}\raggedleft
n\_itr
\end{minipage} & \begin{minipage}[b]{\linewidth}\raggedleft
n\_gc
\end{minipage} & \begin{minipage}[b]{\linewidth}\raggedleft
total\_time
\end{minipage} \\
\midrule\noalign{}
\endhead
\bottomrule\noalign{}
\endlastfoot
tidychangepoint & 6.83ms & 7.16ms & 137.0154 & 73.4KB & 0 & 10 & 0 &
72.98ms \\
changepoint & 857.79us & 878.45us & 1124.8778 & 35.7KB & 0 & 10 & 0 &
8.89ms \\

\end{longtable}

As noted in Section~\ref{sec-coen}, {tidychangepoint} provides two
alternative functions to \texttt{GA::gabin\_Population()} for creating
an initial population for a genetic algorithm:
\texttt{build\_gabin\_population()} and
\texttt{log\_gabin\_population()}. Both perform some näive analysis to
set the probability of an observation being selected uniformly at random
to belong to the first generation. The former runs three quick
algorithms (PELT, Binary Segmentation, and WBS) and sets the probability
to three times the average size of the changepoint sets returned by
those algorithms, divided by \(n\). The latter sets the probability to
\(\log{n}/n\). Both functions can be thought of as ``informed'' by the
data, albeit in different ways. Recall that
\texttt{GA::gabin\_Population()} always uses the probability \(1/2\),
and can thus be considered ``uninformed''.

\begin{figure}[H]

\centering{

\pandocbounded{\includegraphics[keepaspectratio]{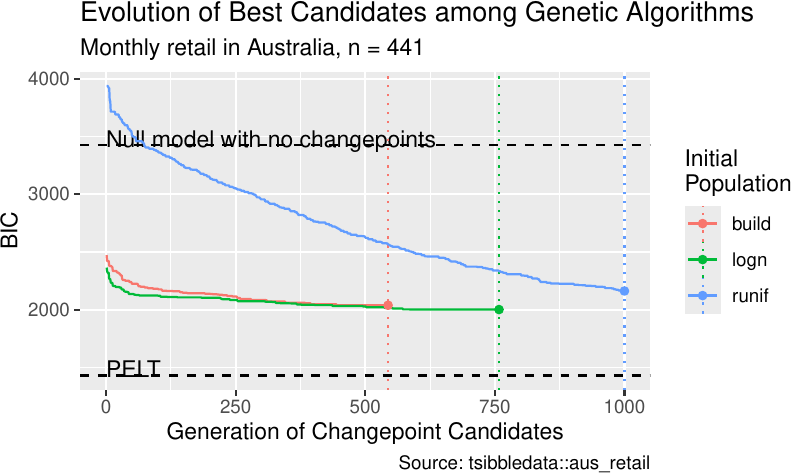}}

}

\caption{\label{fig-compare}Comparison the evolution of the best
candidate changepoint sets for genetic algorithms with different initial
populations. The `runif' series begins by selecting observations to
belong to the candidate changepoint set uniformly at random, with
probability 0.5. On the other hand, the `build' and `logn' series also
selects candidate changepoints uniformly at random, but with informed
probabilities.}

\end{figure}

In Figure~\ref{fig-compare}, we compare the performance of a genetic
algorithm on monthly retail data from Australia (O'Hara-Wild et al.
2022), using these three methods of forming the initial population. In
this test, the maximum number of iterations was set to 1000, with a
consecutive run of 100 generations with no improvement stopping the
algorithm. We see in Figure~\ref{fig-compare} that both of the
``informed'' initial populations result in improved performance. The
recovered changepoint sets score better values of the penalized
objective function, and both algorithms terminated well before they
reached the maximum number of generations. Conversely, the algorithm
using the default initial population was still going after 1000
generations. In fact, it only took 35 generations for the BIC scores
returned by the genetic algorithm whose initial population was set by
\texttt{log\_gabin\_population()} to beat the best score returned by the
same algorithm using the default initial population after 1000
generations.

\phantomsection\label{cell-fig-compare}

Of course, it is possible that by casting a wider net, the default
initial population will catch changepoint sets that were randomly
excluded by the ``informed'' initial populations. But these results held
up under repeated testing, and they suggest a promising line of attack
for future work.

\section{Conclusion}\label{sec-conclusion}

\subsection{Future work}\label{future-work}

There is more work to be done to bring additional changepoint detection
algorithms, models, and penalty functions into {tidychangepoint}. Please
see \href{https://github.com/beanumber/tidychangepoint/issues}{the
package's GitHub Issues page} for a complete list.

Generally, computational costs for changepoint detection algorithms are
a concern. Some algorithms have a known theoretical running time in
relation to the length of the time series, while others do not.
Moreover, while the implementations of some algorithms (notably,
{changepoint} and {GA}) use pre-compiled C code, others may not. All of
the code for {tidychangepoint} is written in R, but since it is mainly
an interface that wraps code from other packages, the overhead is slight
(see Section~\ref{sec-benchmarking}), and there isn't much that can be
done to improve performance. Even the new algorithms that are provided
by {tidychangepoint} (e.g., Coen's algorithm, Shi's algorithm) are
simply special cases of genetic algorithms that use {GA} as a backend.

Another major issue with changepoint detection algorithms is the
robustness of the results to a variety of changes in conditions,
assumptions, or computational environments. While some algorithms report
the probability of an observation being a changepoint, others do not.
Some models rely on assumptions about the distribution or frequency of
changepoints relative to the length of the time series. In randomized
algorithms (such as genetic algorithms), the robustness of the results
may be interrelated with stopping criteria, the initial seed or
population, or other computational considerations. CUSUM methods operate
on a summary of the original time series: are they less susceptible to
outliers? These issues are not correctable within {tidychangepoint}. We
will continue to engage with the authors of the underlying packages when
we see areas for improvement and alert our own users to inconsistencies
through our documentation.

Finally, in this paper we have considered the multiple changepoint
detection problem for a univariate time series, but multivariate time
series are also being studied in a variety of contexts.

\subsection{Summary}\label{summary}

We have described how {tidychangepoint} offers a flexible but
standardized architecture for changepoint detection analysis in R in a
manner that is consistent with the design principles of the {tidyverse}.
Previously, changepoint detection analysis could be conducted using a
variety of R packages, each of which employed its own non-conforming
interface. Our unified interface allows users to compare results across
different changepoint algorithms, models, and penalty functions with
ease.

\section*{References}\label{references}
\addcontentsline{toc}{section}{References}

\phantomsection\label{refs}

\section{Appendix}\label{appendix}

\end{document}